\documentclass[traditabstract]{aa} 
\usepackage[varg]{txfonts}
\usepackage{time}
\usepackage{color}
\usepackage{natbib}
\usepackage[percent]{overpic}
\usepackage{amstext,amsmath,mathtools}
\usepackage{graphicx}
\usepackage{textcomp}
\usepackage{txfonts}
\usepackage{url}
\usepackage{booktabs} 
\begin{document}
\titlerunning{Highly performing solar FPI systems}
\title{The making of robust and highly performing imaging spectropolarimeters for large solar telescopes}

\author{G.B. Scharmer\inst{1}
\and
J. de la Cruz Rodr\'iguez \inst{1}
\and 
J. Leenaarts\inst{1}
\and 
B. Lindberg\inst{2}
\and
P. S\"utterlin\inst{1}
 \and 
Tomas Hillberg\inst{1}
\and
C. Pietraszewski\inst{3}
\and 
A.G. de Wijn\inst{4}
\and
M. Foster\inst{5}
\and
J. Storey\inst{5}}

\institute{Institute for Solar Physics, Dept. of Astronomy, Stockholm University,
AlbaNova University Center, SE 106\,91 Stockholm, Sweden \and
Lens Tech AB, Tallbackagatan 11, SE 931\,64 Skellefte\aa, Sweden \and
IC Optical Systems, 190-192 Ravenscroft Road, Beckenham, Kent, BR3 4TW, United Kingdom \and
High Altitude Observatory, National Center for Atmospheric Research, P.O. Box 3000, Boulder, CO 80307, USA \and
IS-Instruments Ltd., 220 Vale Road, Tonbridge, KENT TN9 1SP, United Kingdom}
\date{Draft: \now\ \today}
\frenchspacing

\abstract{We discuss the requirements, concepts, simulations, implementation, and calibration of two dual Fabry-Perot  (FPI) based imaging spectropolarimeters, CRISP and CHROMIS, at the Swedish 1-meter Solar Telescope, and CRISP2 that is under construction. These instruments are optimised for a large field of view and high cadence at the expense of a moderate spectral resolution, and use a combination of a high-resolution and a low-resolution etalon together with an order-sorting prefilter to define the bandpass. The overall design is made robust and stable by tailoring the low-resolution etalon reflectivity to accommodate expected cavity errors from both etalons, by using a compact optical design that eliminates the need for folding mirrors, and enclosing the entire system within a single container sealed by lenses. By using a telecentric design based on lenses rather than mirrors, image degradation by the FPI system is negligible, as shown in a previous publication, and the throughput of the system is maximised. Initial alignment, and maintaining that alignment over time, is greatly simplified. Moreover, the telecentric design allows full calibration and/or modelling of essential system parameters to be carried out without interfering with the optical setup or the cameras. We also discuss briefly the polarimeters developed for CRISP and CHROMIS. The high performance of CRISP and CHROMIS has been demonstrated in an earlier publication through measurements of the granulation contrast and comparisons with similar measurements simultaneously made through broadband continuum filters. Here, we focus on the aspects of the design that are central to enabling high performance and robustness, but also discuss the calibration and processing of the data, and use a few examples of processed data to demonstrate the achievable image and data quality.  We put forward a proposal for a similar conceptual design for the European Solar Telescope and conclude by discussing potential problems of the proposed approach to designs of this type. Some aspects of these FPI systems may be of interest also outside the solar community.
}

\maketitle

\section{Introduction}
    The construction of 4-m class solar telescopes - the Daniel K. Inouye Solar Telescope \citep[DKIST;][]{2018SPIE10700E..0VW,2020SoPh..295..172R} and the planned European Solar Telescope \citep[EST;][]{2022A&A...666A..21Q} - represents major investments in solar research infrastructures that are motivated in roughly equal parts by the need to resolve the fundamental scales in the solar atmosphere, and the need to improve signal to noise in spectropolarimetry. A major science goal is to observe energetic and dynamic transient events and their associated relatively weak magnetic fields in the chromosphere and corona at the highest possible spatial resolution, and with a cadence that matches their temporal evolution. In addition to these requirements, which already pose major challenges, there is the need to observe the large-scale structuring of magnetic field in active regions. Such studies, which ideally cover more than three decades in spatial scales, ultimately aim at  explaining small- and large-scale energetic events, such as flares and coronal mass ejections (CMEs), in the complex context of large-scale changes and instabilities of the magnetic field in active regions.

The above science goals could be addressed through the development of a fictitious highly transmitting spectropolarimeter that covers a large field of view (FOV) with high spatial and spectral resolution at many wavelengths, with a single exposure. Presently, there is no instrument that can meet all these requirements. Conventional slit spectrographs are not of interest in this context - the need to obtain spatial coverage by scanning implies that either the spatial coverage has to be extremely limited or cadence totally sacrificed. Developing efficient so-called Integral Field Spectropolarimeters is a longstanding goal aiming to overcome the shortcomings of conventional slit spectropolarimeters \citep[for overviews of past and recent efforts, the reader is referred to][ and references therein]{2000SPIE.4008.1181D,2013JAI.....250009C,2022A&A...668A.149V}. 

One approach to the design of Integral Field Spectropolarimeters is through an image slicer, such as that installed at GRIS on GREGOR \citep{2012AN....333..872C,2014SPIE.9147E..3IC} - a similar design has been proposed for EST \citep{2013JAI.....250009C}. However, the number of equivalent spectrograph slits sampling spectra in parallel is in practice limited to be on the order of ten or a few tens, which defines the spatial coverage possible with a single exposure. An even more efficient implementation is made possible with the development of micro-lens based spectrometers, in which the slit is replaced with a 2D array of microlenses, each of which creates a short spectrum. The bandwidth of this spectrum, which is re-imaged on the detector, is set by the passband of a prefilter. Proposals for such microlens based spectrographs were made already  several decades ago \citep{1988igbo.conf..266C,1994SPIE.2198..229O,1995ASPC...71..239B}. A microlens based integral field unit (MEIFUS) was later proposed for 8-m and EELT-class night-time telescopes the by \citet{2003SPIE.4842..174C}. A microlens based solar spectropolarimeter has been successfully implemented on the TRIPPEL spectrograph \citep{2011A&A...535A..14K} at the Swedish 1-m Solar Telescope \citep[SST; ][]{2003SPIE.4853..341S} by \citet{2022A&A...668A.149V} and \citet{2022A&A...668A.150V}, and is referred to as MiHi$^2$. An even more ambitious implementation is that of HeSP \citep{2025A&A...696A...3L}, which targets the He I 1083 nm line, formed in the uppermost chromosphere. The benefit of the microlens approach is the relative ease by which many microlenses can be used to reimage multi-wavelength data from a cohesive FOV that is large enough to allow the use of image restoration techniques to compensate for seeing-induced image degradation. However, a hard upper limit to that FOV is set by the size and number of pixels of the detectors, because of the relatively large number of spectral pixels needed for each spatial sample. In practice, this means that a 10"x10" FOV is extremely difficult to reach with a microlens based instrument on a 4-m class telescope. Even on the 1-m SST, the FOV of MiHi$^2$ and HeSP is only 8"$\times$8".

Because of the aforementioned problems in covering a large FOV with a single exposure, using either a spectrograph based image slicer or microlens array replacing the slit, imaging spectropolarimeters based on Fabry-Perot Interferometers (FPIs) remain of considerable interest  \citep{1998A&A...340..569K, 2006SoPh..236..415C, 2008ApJ...689L..69S, 2011SoPh..268...57M, 2012AN....333..880P, 2012SPIE.8446E..77K, 2014SPIE.9147E..0ES, 2020A&A...642A..11S, 2017psio.confE..85S, 2022PASP..134a5007G, 2025arXiv250208268D}. Such instruments can be constructed to cover a relatively large (about 1 arcmin) diameter FOV on a 4-m class telescope, at the expense of the necessity of sampling spectral lines one wavelength at a time by tuning the FPIs. For solar applications, the cadence needed to follow dynamic and energetic events, combined with high requirements on signal to noise, constrains the possible number of wavelength samples to such an extent that it becomes contra productive to build an imaging spectropolarimeter with high spectral resolution. A related problem is that the time needed to complete an individual spectral scan can distort the interpretation of dynamic events \citep{2018A&A...614A..73F}. Noting also that spectrally undersampling data leads to increased errors in measured field strengths and line of sight velocities \citep[][Fig. 3]{2017SSRv..210..109D}, low or medium spectral resolution (of about 50-100,000) is strongly favoured. This makes it possible to scan most spectral lines at a relatively small number (5-10) wavelengths samples, though many more samples may be needed to cover very strong lines, such as the Ca~II H and K lines.

To enhance the efficiency of FPIs, especially in the context of diagnostics relating to the chromosphere, it is an obvious advantage to be able to operate two or more FPIs in parallel, each covering its own wavelength range. At SST, CRISP \citep{2008ApJ...689L..69S}, soon to be replaced with CRISP2,  covers the wavelength range 500-860~nm, whereas CHROMIS \citep{2017psio.confE..85S} covers the wavelength range 390-500~nm. For EST, it is foreseen to use three FPIs operating in parallel, covering the wavelength ranges 380-500~nm, 500-680nm, and 680-1000~nm  \citep{2022A&A...666A..21Q}. 

In this paper, we describe the requirements, concepts, simulations, implementation, and calibration of  the imaging spectropolarimeters used at SST and proposed for EST. 
To understand the requirements and the design, we explain in Sect. \ref{insights_decisions} how science data are obtained with an imaging spectropolarimeter, which sequentially tunes through one or more spectral lines and samples their different polarisation states. We then review the decisions, calculations and simulations of \citet{2006A&A...447.1111S}, which led to the design of CRISP and later CHROMIS and CRISP2, which are compact, robust, cost efficient and highly performing imaging spectropolarimeters covering the wavelength range 380-860~nm. In Sect. \ref{image_restoration}, we explain the processing and calibration of the data, and in Sect. \ref{CRISP_performance} we discuss the performance of CRISP and CHROMIS and present sample data to illustrate the achievable data quality. Section \ref{Conceptual_designs} discusses the conceptual designs of seven FPI systems for SST and EST, with emphasis on three of them, and Sect. 6 summarises the choices and designs of cameras and polarimeters for CRISP and CHROMIS. Section \ref{conclusions}, finally, summarises the main results and discussions. 
 
\section{Insights and decisions leading to the designs of CRISP and CHROMIS} \label{insights_decisions}
 \subsection{Basic operation and challenges}

 An FPI-based imaging spectropolarimeter typically operates by sequentially tuning through polarimeter states (in the innermost loop), wavelengths across one or more spectral lines (middle loop), and a repetition loop (outermost loop). The overwhelming challenge of such a system operating on a groundbased solar telescope is the variability of seeing, which rarely is stable on the time scales needed to complete even a single full cycle of measurements. The variability of the point spread function (PSF) during such observations gives rise to serious cross-talk that may seem prohibitive for any meaningful analysis of the data. However, we have repeatedly demonstrated the possibility of using CRISP and CHROMIS to provide multi-wavelength diagnostics of the solar atmosphere at a spatial resolution close to the diffraction limit of the telescope by using short exposures (typically 10-20 msec) and advanced image restoration techniques \citep{1994A&AS..107..243L, 2002SPIE.4792..146L, 2005SoPh..228..191V, 2021A&A...653A..68L} that compensate for much of the variability of the seeing,  \citep{2019ApJ...870...88E, 2019A&A...621A..35L, 2020A&A...637A...1K, 2021A&A...647A.188D, 2021A&A...649A.106Y, 2021A&A...650A..71K, 2022A&A...664A...8M, 2022A&A...661A..59D, 2022A&A...664A..72J, 2023A&A...672A.141E, 2024A&A...683A.190R, 2024A&A...686A.218N, 2024A&A...691A.198J,  2024A&A...688A..56B, 2025A&A...696A.105S, 2025A&A...696A.125P, 2025A&A...693A.165F}. Furthermore, we have demonstrated that such data can be used to infer the detailed dynamic and magnetic properties of the atmosphere. Below, we first explain the critical decisions and insights behind the successful designs of CRISP and CHROMIS. 
 
\subsection{Telecentric vs collimated etalon mounts} \label{telecentric_vs_collimated}

Early in the project, it was decided that the reimaging system of CRISP should be telecentric, i.e., the etalons are placed close to a focal plane, and the pupil is reimaged at infinity. The alternative is to locate the etalons close to an optical conjugate of the telescope pupil with the image at infinity, which is referred to as a collimated mount. The latter arrangement is attractive, because in principle it delivers a spectral transmission profile that is the same everywhere but that is shifted in wavelength by an amount that varies as the square of the field angle. However, there are two caveats that disfavour the use of collimated mounts. The most important one comes from the need to suppress or eliminate ghost images and transient interference fringes caused by multiple inter etalon reflections. In general, ghost images are a major concern because of the high reflectivities of the coated surfaces of the etalons. One option proposed previously is to locate the prefilters of the FPI system between the two etalons. However, this requires prefilters that are of the same size as the etalons, or complicated intermediate re-imaging systems to decrease their needed diameter \citep{2011SPIE.8172E..19G}. Furthermore, the suppression of ghost images with this technique is inadequate if the transmission of the prefilter is high, which is most often the case of modern interference filters. The only effective method of suppressing the ghost images that we are aware of is by tilting one of the etalons relative to the other, but then the symmetry of the system is broken and the shape of the transmission profile will change across the FOV. This defeats the main advantage of the collimated mount. 

The other (and less decisive) argument against collimated mounts is image quality. Random cavity errors of etalons are amplified by multiple reflections in the cavity, and this can produce accumulated phase errors that significantly degrade image quality if the reflectivity is high \citep{2000A&AS..146..499V, 2006A&A...447.1111S}. In addition, any significant variations of the transmission across the pupil, caused by the same random cavity errors, will lead to a PSF that varies strongly with wavelength across the combined passband of the two etalons (this problem occurs also in a single etalon system). Calibrating and compensating for this PSF by mapping cavity errors and phase errors across the pupil and at many wavelengths across the passband of a dual FPI system appears extremely complicated and unlikely to be part of any standard data reduction procedure. Furthermore, small changes in the co-tuning of the etalons could change the PSF significantly. Nonetheless, FPI systems with small etalons of outstanding quality have indeed been built and have demonstrated the feasibility of collimated mounts with high image quality \citep{2006SoPh..236..415C}. 
 
With a telecentric mount, the etalons are located close to a focal plane, which is safe in terms of image quality. In CRISP and CHROMIS, the FPIs are located at a distance from the focal plane chosen such that the PSF is smeared to a diameter of about 1 mm. This has the consequence that only cavity errors at scales below 1~mm can degrade image quality, which is unlikely to cause noticeable effects. However, cavity errors of the two etalons will cause random changes of their combined spectral transmission profile across the FOV, which can have a huge impact on the quality and interpretations of the data. These spectral transmission profiles are characterised by four parameters, pixel by pixel over the FOV, namely the reflectivities and cavity errors of the two etalons. The good news is that these parameters can be measured using standard calibration procedures, which are made on a daily basis at SST (Sect. \ref{mapping_etalons}), such that the spectral transmission profile can be considered known in each camera pixel. Of particular importance is that this calibration can be made without disturbing the optical setup at all. The bad news is that cavity errors can cause this transmission profile to vary wildly over the FOV, and even show double-peaked transmission profiles if the reflectivities of the etalons are high. Controlling the shape of the transmission profile therefore is a high priority in the design of a robust dual FPI system, and how to do this is explained in Sect. \ref{simulation_insights}.

Building large FPI systems with high requirements, such as those for the VTF on DKIST  \citep{2012SPIE.8446E..77K, 2014SPIE.9147E..0ES, 2024SPIE13096E..17H} and the presently proposed FPI systems for EST \citep{2025arXiv250521053S}, represents more challenges than for existing smaller systems. One challenge is that the overall size of the FPI system increases with increasing telescope aperture diameter. Limitations in available space then imposes constraints on the optical design. For VTF, the solution was to fold the optical beam multiple times. For EST, we have chosen instead to make the system as compact as possible without folding the beam, building on the successful designs of CRISP and CHROMIS. This represents the biggest difference between the the DKIST and EST  FPI systems. Another challenge is the difficulty of manufacturing large etalons with sufficiently small cavity errors. The EST FPI systems use smaller etalon diameters than those of VTF, which should help in (reducing the costs of) maintaining cavity errors at an acceptable level. VTF and the FPI systems for EST are all telecentric, which according to the previous discussion is safer than using a collimated design, and have similar spectral resolutions at comparable wavelengths.

\subsection{Insights gained from numerical simulations}\label{simulation_insights}

The design of CRISP, and later CHROMIS, relies on two important insights gained from simulated behaviour of dual etalon performance \citep{2006A&A...447.1111S}, when used with a telecentric mount. The first insight is that the damaging effects of cavity errors on dual etalon performance can be mitigated by combining a high resolution, high reflectivity etalon with a low resolution etalon that has much lower reflectivity. The lower reflectivity of the low-resolution etalon increases the width of its passband to more easily accommodate the random relative wavelength shifts of the peak transmissions of the two etalons, caused by the unavoidable cavity errors. By properly using the reflectivity of the low resolution etalon as a free parameter to balance the impact of the cavity errors of the etalons, the shape of the spectral transmission profile can be controlled to show only small variations over the FOV. This simple trick allows a dual etalon system to operate with high transmission and relatively small variations in the shape of the transmission profile over the FOV, even when cavity errors are large. For large etalons, such as required by EST, this is of particular importance, since increasing the clear aperture makes small cavity errors increasingly difficult to reach. This robustness also facilitates precise measurements of the transmission profiles in each pixel, to augment the interpretation of the data. A comment here is that having a variable transmission profile across the FOV may seem awkward to an optical engineer, and may also raise the suspicion that taking into account these variations will cause substantial overhead while processing data from the FPI system. However, the (polarised) radiative transfer calculations employed with sophisticated inversion techniques are sufficiently advanced that this overhead is negligible. Furthermore, by taking into account the actual shape of the transmission profile, the inversion results can be made robust and independent of variations in the transmission profile.

 The second insight gained from our simulations \citep{2006A&A...447.1111S} is that the damaging effects of phase error on the image quality for FPIs in telecentric mounts, claimed by \citet{2000A&AS..146..499V} to a large extent can be compensated for by refocusing \citep{2006A&A...447.1111S}, later confirmed by \citet{2010A&A...515A..85R}.  Thus, pupil apodisation, originating from the more tilted rays of the outer parts of the pupil in a telecentric reimaging system having FPI transmission peaks that are blue-shifted relative to those from the center of the same pupil, remains the dominant effect limiting image quality, as first proposed by \citet{1998A&AS..129..191B}. The consequence of this is that telecentric FPI systems can be built with a focal ratio at the etalons that can be much smaller than indicated by the calculations of \citet{2000A&AS..146..499V}. Unfortunately, this still requires a telecentric FPI system to be used in a very slow beam, and thus to have a very large clear aperture when used with a large telescope.

\subsection{Optical design aspects} \label{design_aspects}

An important factor in explaining the successful design of CRISP is the decision to make the overall optical system as compact as possible, and base it on lenses without any folding mirrors. This decision was based on the reasonable conjecture that a straight-through optical system without mirrors would simplify initial alignment, and maintaining that alignment over time. Moreover, with adequate anti-reflection coatings on the lens surfaces, the overall throughput can be made higher than with mirrors. The installation and commissioning of CRISP and CHROMIS support our conjecture: in both cases, installation took only a few days and the first high quality data were recorded immediately thereafter.  In the case of CRISP, the size of the existing optical table and the  main observing room of SST constrained the overall length of the system to be less than about 1.5~m, whereas CHROMIS could be allowed to be somewhat longer. In both cases, it was possible to design FPI systems operating at high Strehl with these constraints on their overall lengths. 

\subsection{Wideband images}\label{wideband_images}
A crucial aspect of the CRISP and CHROMIS FPI systems is its close association with a dedicated auxiliary wideband system that uses light from a broad interference filter centered on (nearly) the same passband as the narrowband system, along with cameras that record synchronised exposures from the two systems. This wideband system is of such critical importance, that the narrowband system would be essentially useless without it.

The wideband system uses a broad interference filter that displays solar photospheric fine structure, or that is strongly dominated by such fine structure. This is in order for the wideband to serve as a stable reference (acting as an anchor channel), which shows exactly the same solar target while the narrowband system tunes through a spectral line and various polarisation states, entailing a wide range of photospheric and/or chromospheric fine structure. This stable wideband reference is crucial for the processing of the data \citep[Sect. \ref{image_restoration};][]{2005SoPh..228..191V}, because any variability of the recorded wideband images directly reflects the variability of the seeing. Since these are recorded at exactly the same time as the narrowband images, and at nearly the same wavelength, the wideband images hold crucial clues to the restoration of the narrowband images. The additional requirement that the fine structure seen through the wideband filter is photospheric is in order to allow the wideband image to serve as a reference for co-aligning multi-line data recorded at widely different wavelengths, or even recorded with entirely separate FPI systems (such as CRISP and CHROMIS). By forcing the wideband images to consistently show the solar photosphere, we can use conventional cross-correlation techniques applied to the wideband images to accurately co-align data sets recorded even at widely different wavelengths. In most cases, the prefilter of the FPI system can be used also as wideband filter but in the case of extremely strong lines, such as the Ca II H and K lines, the FPI prefilter is not wide enough to show photospheric structure. In that case, a wideband filter centered on a nearby (quasi-)continuum is used instead.

\section{Image restoration and calibrations}\label{image_restoration}
\subsection{Image restoration}
As discussed already, the variability of daytime seeing, even when using adaptive optics at the best sites in the world, is a major challenge in obtaining high-quality multi-wavelength polarimetric data sets with a ground based solar telescope. The key to compensating this variability is through the use of advanced image restoration techniques. Such techniques have been developed with the purpose of improving data from the 50-cm Swedish Vacuum Solar Telescope (SVST, Scharmer et al. 1986) and SST, starting with Löfdahl and Scharmer (1994). The first of these techniques aimed at restoring pairs of focused and defocused (phase diversity) images recorded at a single wavelength. Since then, these techniques have matured and been extended into the processing of multi-wavelength, multi-polarisation state data sets, with and without a defocused channel  \citep{2002SPIE.4792..146L, 2005SoPh..228..191V, 2015A&A...573A..40D} . 

Of particular importance is that a defocused channel is not necessary for successful restoration of degraded images - even a sufficiently large set of focused images, assuming wavefront diversity originating from seeing, contain information about the aberrations that cause the image degradation. This is because a spatial distribution of phase errors across the telescope pupil corresponds to numerous interferometer pairs at any particular spatial separation (which are mapped into a corresponding spatial frequency), which are partly out of phase with each other. This decreases the total response (amplitude of the optical transfer function - OTF) at that spatial frequency, such that the processing algorithm will recognise a drop in the OTF amplitude as necessarily originating from a spatial phase error. By modelling the variation of the phase across the pupil, the algorithm can make a reasonable estimate of the distribution of phase errors, and compensate the restored image accordingly. Note that the algorithm may actually fail to reproduce the detailed spatial variation of the phase errors at the pupil, but it will capture more correctly its impact on the OTF and PSF. The technique of restoring focused images without a defocused pair is referred to as "multi-frame blind deconvolution" and is part of a complex and more general scheme of processing complex data sets later coined multi-object multi-frame blind deconvolution (MOMFBD) techniques \citep{2002SPIE.4792..146L}. The particular implementation of these techniques was first developed by \citet{2005SoPh..228..191V} and later used with CRISP data, with an updated methodology and data processing pipeline specifically targeting the processing of CRISP and CHROMIS data \citep{2015A&A...573A..40D, 2021A&A...653A..68L}. 

\begin{figure}
\center
\includegraphics[angle=0, width=\linewidth,clip]{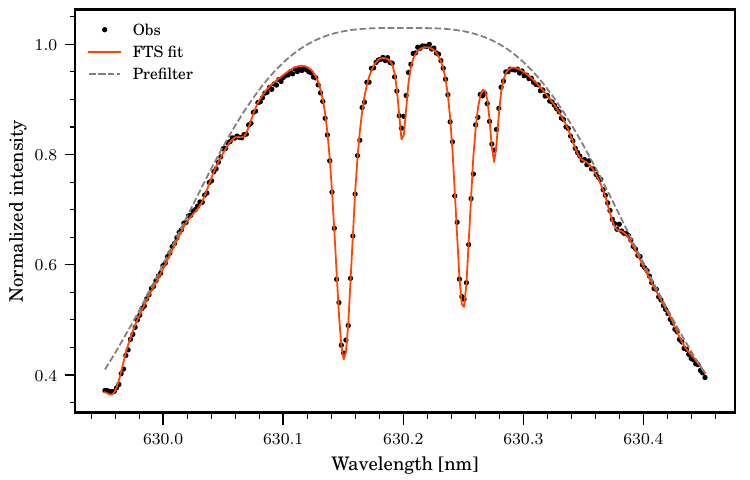}
 \caption{
Averaged scans of a spectral region of the solar Fe~I lines and the telluric lines around  630.2~nm recorded with CRISP (dots) together with the scanned FTS spectrum of \citet{1984SoPh...90..205N}, degraded by the transmission profile of CRISP, as determined from fits of the reflectivities and cavity errors shown in Fig. \ref{fig:etalons}. The dashed curve corresponds to the transmission curve of the prefilter, as obtained from the same data.}
\label{fig:prefilter}
\end{figure}

\begin{figure*}[!h]
\centering
\includegraphics[angle=0, width=0.77\textwidth,clip]{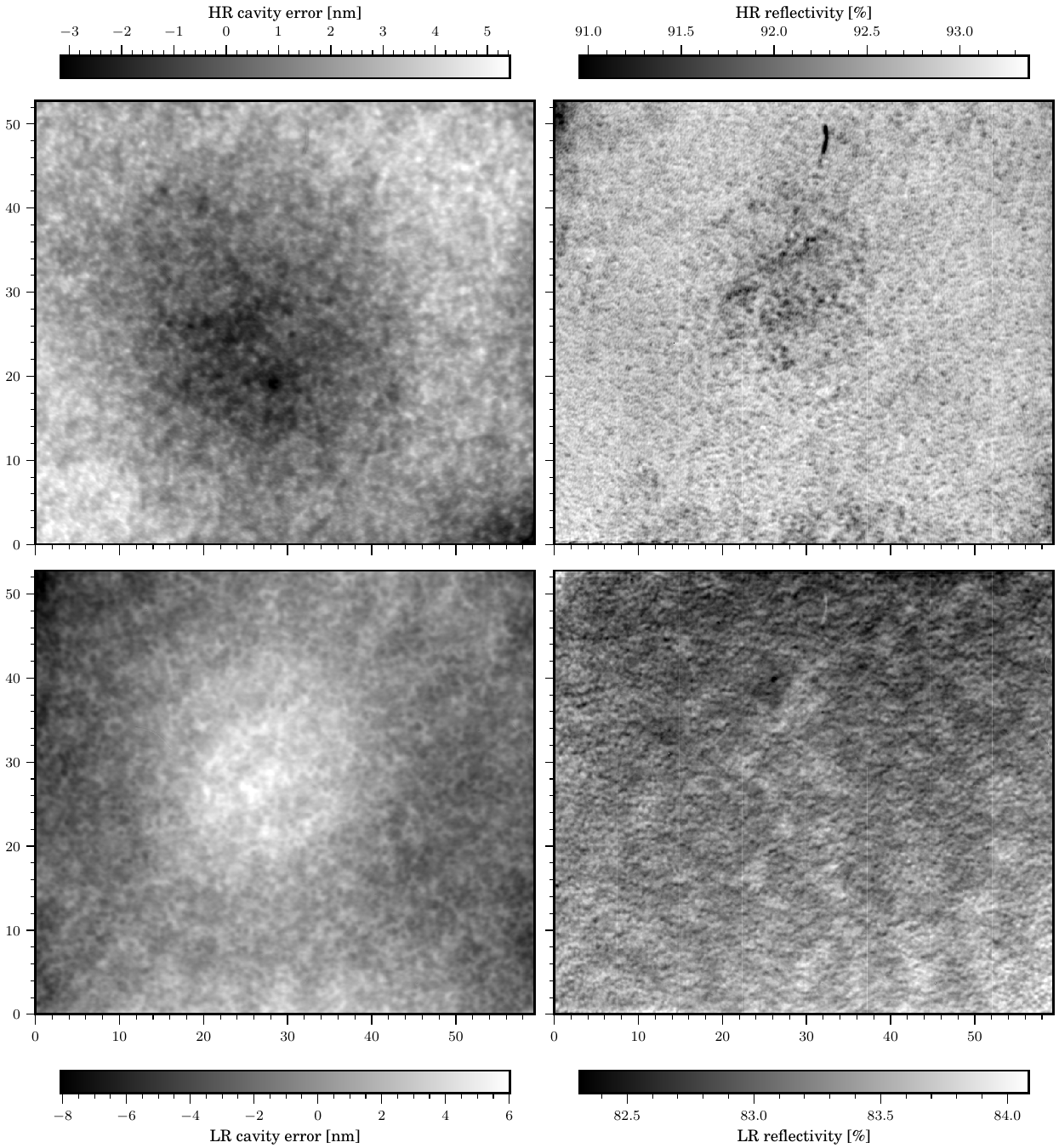}
 \caption{
Cavity maps and reflectivity maps of the high-resolution (HR) and low-resolution (LR) etalons of CRISP, obtained by first recording images while scanning the two etalons together, then scanning the LR etalon across the passband of the HR etalon tuned to a continuum wavelength, and finally fitting the scanned profiles from the Fourier Transform Spectrometer at the McMath-Pierce Telescope \citep{1984SoPh...90..205N}.
}
\label{fig:etalons}
\end{figure*}

The key to the processing of the FPI data is the synchronised exposures with the wideband system. Whether or not the wideband system delivers also defocused images (which is not always the case with CRISP and CHROMIS), the wideband images contain crucial information about the degradation of the PSF for each pair of narrowband and wideband exposures constituting a data set. Such a data set is composed of typically several hundred images, recorded during 10-15~sec, corresponds to narrowband images scanned in wavelength, covers four polarisation states, and with scans repeated several times to improve noise in the data. These images and the corresponding simultaneously exposed wideband images are processed as a single data set, constrained by the following rules:
\begin{itemize}
\item All wideband images correspond to a single unknown object. 
\item All narrowband images recorded at a particular wavelength and at a particular polarisation state corresponds to one unknown object. There are $n_\lambda \times n_p$ such unknown objects, where $n_\lambda$ is the number of wavelengths scanned, and $n_p$ the number of polarisation states.
\item The aberrations are unique for each exposed wideband image but identically the same for the simultaneously exposed narrowband image.
\end{itemize}  

For the above technique to work well, the FOV is segmented into subfields that are typically a few arc seconds wide. This is necessary in order to compensate also for high-altitude seeing, in particular for that arising in the layers corresponding to the jet streams around the tropopause, in addition to the seeing originating from near ground. Image restoration is performed independently for these subfields but with spatial overlaps such that interpolation can be used to provide a smooth transition between the restored subfields. A possibly better approach could be to model the atmosphere as a small number of discrete seeing layers, which would connect the subfields to each other by their geometry and thus decrease the number of free parameters of the image restorations. Such a development is in progress (Löfdahl, private communication). 

\subsection{Calibrations}
\subsubsection{Alignment between the narrowband and wideband systems}
To ensure the correspondence between the seeing variations of the narrowband and wideband images and to accurately co-align the wideband and narrowband images, it is necessary to calibrate their relative alignment, pixel by pixel. At SST, this is done by inserting a pinhole grid at the Schupmann focal plane just below the vacuum system. A beam splitter cube just in front of the narrowband FPI system splits off 5-10\% light to the wideband system and allows both systems to simultaneously record the pinhole pattern. This procedure is repeated, as part of a standard calibration procedure, every day. In case an intermediate focal plane similar to the Schupmann system at SST is not available, it is possible to use a focal plane just in front of the FPI system for the location of the pinhole grid, but then the beam splitter sending light to the wideband system must be located between that focal plane and the first lens of the FPI system, and a similar calibration must be made for each FPI system independently. 

It is important that the optical setup is stable such that the calibration of the relative position between the wideband and narrowband images can be applied to science data recorded several hours earlier or later. Note that there are two levels of accuracy required here: one is to ensure that the seeing is the same for the wideband and narrowband images - this requirement can probably be relaxed to a few tenths of an arc second. However, the wideband images are almost always used as part of the science data set with the narrowband images, directly or indirectly (see below), and that requires a stability of the alignment for at least five hours at the level of a camera pixel, or even a small fraction thereof. 

Note that the above procedure is not applicable for co-aligning observations made at (widely) different wavelengths, or with separate FPI systems. For co-alignment of such data, differential atmospheric refraction, which varies strongly with solar inclination and thus with time, invalidates a pinhole grid calibration. By consistently requiring the wideband images to show photospheric fine structure, we can co-align the wideband images recorded at different wavelengths using conventional cross-correlation techniques, and through their individual pinhole calibrations align also narrowband data sets recorded at widely different wavelengths. Even if the wideband images are not used directly as part of any science data set, they play a fundamental role in connecting narrowband data recorded at (widely) different wavelengths. Therefore, the alignment of such multi-wavelength narrowband data requires stability between the narrowband FPI systems and their wideband counterparts at the levels of a fraction of a pixel over time scales of several hours.

The accuracy and stability of alignment calibrations made with pinhole grid patterns thus are crucial in enabling a high level of fidelity of complex science data sets obtained with FPI systems. \citet{2008A&A...489..429V} found that the typical accuracy in the alignment of MOMFBD restored objects based on pinhole grid calibrations is about 0.05~pixel, and that such accuracy is needed for the alignment of the images from the two channels of a polarising beam splitter to prevent polarisation artefacts. This is of particular importance when measurements of polarisation signals with very low noise levels are aimed for. Somewhat less demanding, but still very challenging, is to accurately align narrowband data recorded at different wavelengths. A better understanding of these critical requirements through simulations would be desirable.

\subsubsection{Mapping cavity errors and reflectivity variations}\label{mapping_etalons}
To map the spatial variation of the  spectral transmission profile of the FPI system, we need to calibrate the two reflectivities and the cavity errors pixel by pixel for every prefilter-defined spectral regions of interest. This is done, without interfering with the optical setup, by recording the average quiet Sun profile of  a suitable spectral line within the passband of the prefilter while random voltages on the adaptive mirror and rapid circular movement of the telescope near disk center blurs the solar granulation pattern. To map the cavity errors and reflectivities of both etalons, two separate scans are made (de la Cruz Rodr\'iguez et al., in prep.). First, the two FPIs are co-tuned while scanning the spectral line. During the second scan, the passband of the high-resolution FPI is fixed on a continuum wavelength away from the spectral line and the low-resolution FPI is scanned in wavelength. Information about the cavity errors comes from the wavelength shifted maxima of the two scanned profiles, and information about their reflectivities comes from the FWHM of the scanned profiles. By fitting the scanned profiles to the highly resolved spectra of the Fourier Transform Spectrometer (FTS) at the McMath-Pierce Telescope \citep{1984SoPh...90..205N}, pixel by pixel, the cavity errors and reflectivities of the two etalons can be calculated with adequate accuracy for determining the spectral transmission profile of the dual FPI system in each pixel. These fits are also used to characterise the variation of the prefilter properties over the FOV, though such fits are not shown here. Figure \ref{fig:prefilter} shows such a scan through the solar and telluric spectral lines around 630.2~nm. The observed profile shown represents the averaged profile over the FOV together with the scanned FTS spectrum, degraded by the obtained spectral transmission profile of CRISP. Note that the telluric lines are not included in the fits, which are based entirely on the solar lines. Figure \ref{fig:etalons} shows the maps of the cavity errors and reflectivities obtained as explained in the caption. Though these calibrations can provide maps over both cavity errors and reflectivity variations over the FOV, the actual variations in reflectivity have only a minor impact on the spectral transmission profile compared to that of the cavity errors, and can be ignored. 

\begin{figure*}[!h]
\centering
\includegraphics[angle=0, width=0.92\linewidth,clip]{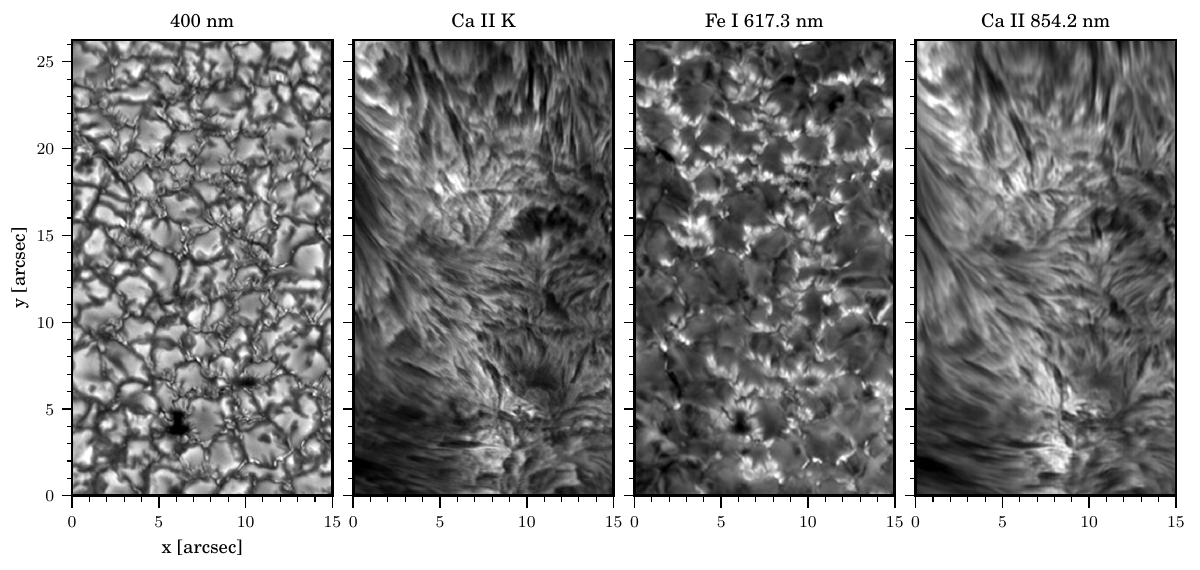}
\includegraphics[angle=0, width=0.92\textwidth,clip]{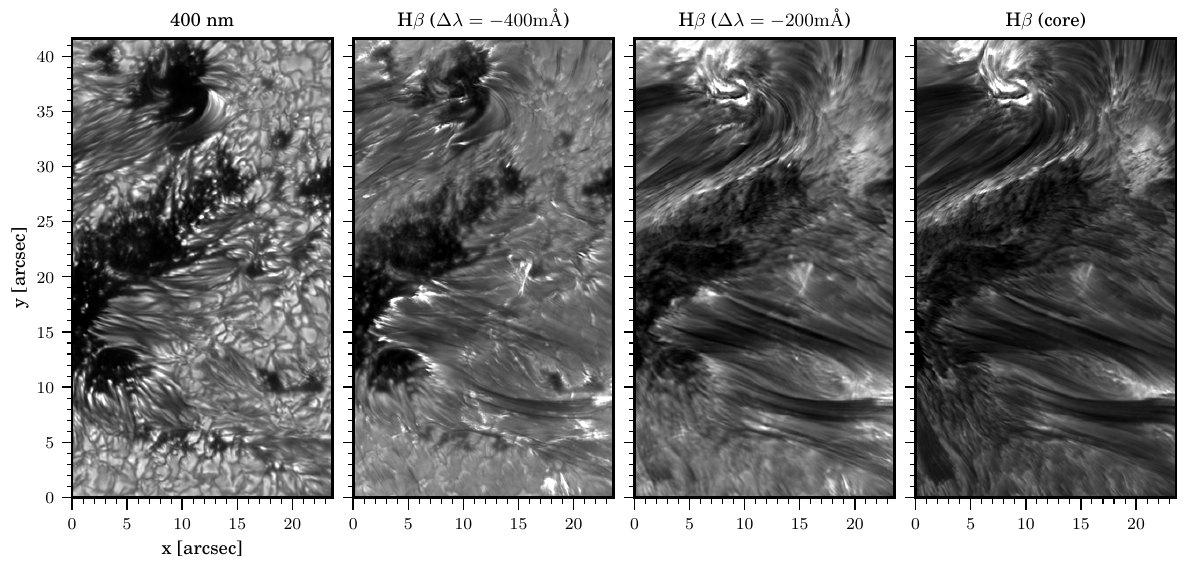}
 \caption{
Collage of high resolution CRISP and CHROMIS images showing a plage (top) and parts of active regions  (bottom). The 400~nm (clean continuum), line core Ca~II K , and H$\beta$ images were recorded with CHROMIS, the Fe~I 617.3~nm and Ca~II 854.2~nm images are line core images recorded with CRISP. These images, selected from multi-wavelength scans through the corresponding spectral lines, illustrate the wide range of multi-line diagnostics possible by combining data from CRISP and CHROMIS, and the quality of the data.}
\label{fig:AR+plage}
\end{figure*}

\subsubsection{Calibration of point spread function parameters}
With a telecentric mounting, the spatial PSF is given by the focal ratio of the beam at the etalons and their reflectivities and cavity separations that vary over the FOV. However, for all designs described here, this focal ratio and the average reflectivities and cavity separations chosen are such that the Strehl degradation by apodisation effects and phase errors are 5\% or less, which is very small compared to the image quality degradation caused by the seeing and residual phase errors left by the AO system. Furthermore, the small variations of the reflectivity and cavity separations over the FOV are of insignificant importance in this context such that the PSF in practice can be regarded as independent of the field position. If needed, the actual PSF can be calculated on the basis of the reflectivities and cavity separations averaged over the FOV. For further details on the calculation of the PSF, see \citet{2006A&A...447.1111S}. 

\section{Performance of CRISP and CHROMIS} \label{CRISP_performance}
The quality of data from CRISP and CHROMIS obviously depends on several factors, such as the seeing quality at the site of SST on La Palma, the design and optical quality of SST \citep{2003SPIE.4853..341S}, and the efficiency of its 85-electrode adaptive optics system \citep{2024A&A...685A..32S}. To understand performance limitations of SST and its adaptive optics, we have combined measurements of the seeing, characterised by the Fried parameter $r_0$ \citep{1966JOSA...56.1372F}, and measurements of the contrast of quiet Sun granulation near disk center, averaged over 2 sec. intervals \citep{2019A&A...626A..55S}. The excellent correlation between the granulation contrast and $r_0$ in Fig. 6 of that paper shows that granulation contrast can serve as an excellent proxy for seeing quality when a large solar telescope is used with a high-order AO system. This is, as is well-known from conventional AO systems \citep[Sect. 4.4.4]{1998aoat.book.....H} , because the residual high-order aberrations that are beyond the capability of the AO system leaves a PSF that has a diffraction limited core combined with a halo, the width of which is given by the diameter of the wavefront sensor subapertures. In the case of SST, the halo will have a diameter of about 1 arc second, which happens to correspond roughly to the scale of the solar granulation pattern, and the Strehl corresponds to the fractional energy of the PSF in the diffraction limited core. The granulation contrast is reduced by the wings of the PSF but not its core, explaining the excellent correlation between $r_0$ and the granulation contrast \citep{2019A&A...626A..55S}.   

To evaluate the quality of CRISP itself, without influence of the above mentioned factors, we have recorded simultaneous images with the wideband and narrowband systems, both using the same prefilter \citep{2019A&A...626A..55S} but with the narrowband system tuned to a continuum wavelength within the passband of that prefilter. Using 3D MHD simulations, we have calculated the expected contrast for both the narrowband and wideband systems. Comparing the measured granulation contrasts measured with the wideband and narrowband systems in a wide range of seeing conditions, the correlation is found to be perfect and nearly exactly with the ratio of contrasts expected from the synthetic spectra obtained from MHD simulations. This suggests that the FPI system does not degrade image quality by any significant amount. We attribute this primarily to the telecentric setup of the etalons and the telecentric re-imaging system. The latter leads to lens diameters that are almost entirely set by the diameter of the FOV and with the pupil projected onto these lenses being only a small fraction of the lens diameters. This effectively means that wavefront errors at large scales, whether it is the lenses or the FPIs,  cannot degrade image quality. Producing lenses with such relaxed tolerances is not challenging.

\begin{figure*}[!h]
\centering
\includegraphics[angle=0, width=0.95\textwidth,clip]{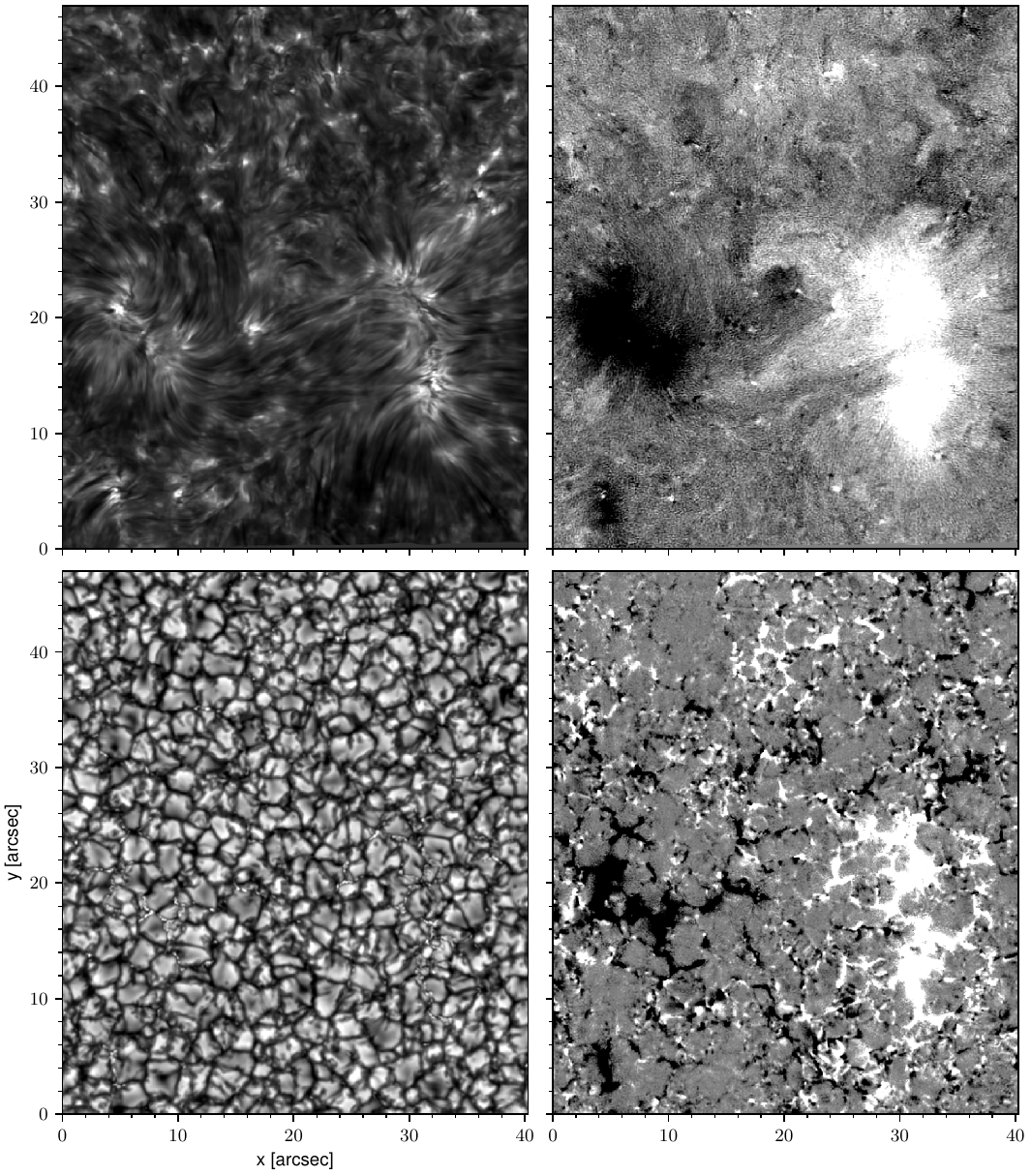}
 \caption{
High resolution CRISP data connecting the solar chromosphere (top) and photosphere (bottom). The top two panels show a line core Ca~II 854.2~nm image of the quiet Sun, and the LOS magnetic field obtained from a polarimetric scan through the same line. The inversion with the Ca~II 854.2~nm data uses the weak-field approximation with spatio-temporal coupling \citep{2024A&A...685A..85D} and the chromospheric LOS magnetic field map shown is clipped outside $\pm$50~G. The lower two panels show an image recorded in the continuum near the Fe~I 617.3~nm line, and the LOS magnetic field obtained from a regularised Milne-Eddington inversion of a polarimetric scan through the same line. The photospheric LOS magnetic field map shown is clipped outside $\pm$25~G.
}
\label{fig:LOS_magnetic_field}
\end{figure*}

\begin{figure*}[!h]
\centering
\includegraphics[angle=0, width=\linewidth,clip]{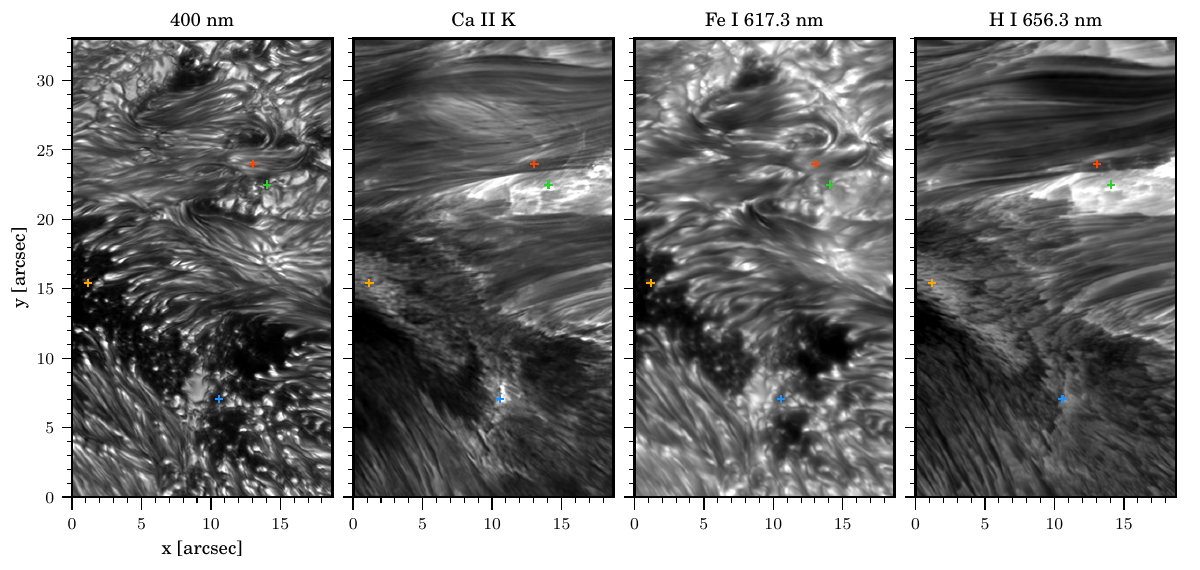}
\includegraphics[angle=0, width=\linewidth,clip]{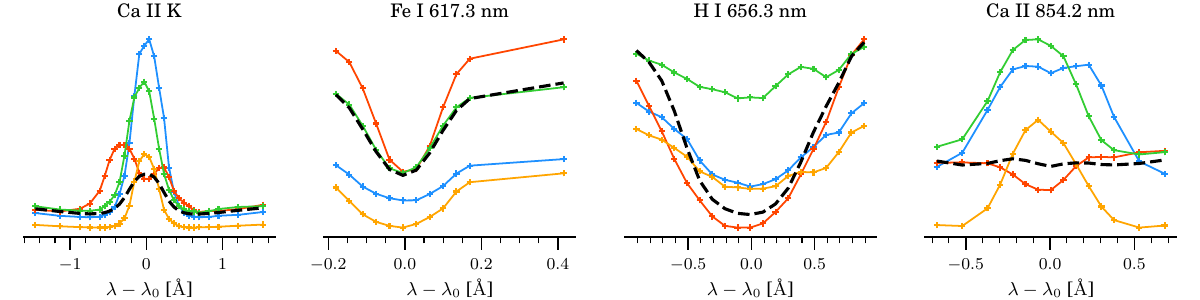}
 \caption{
High resolution CHROMIS (400~nm and Ca~II K) and CRISP (617.3~nm and H$\alpha$ 656.3~nm) images displaying the photospheric and chromospheric fine structure of an active region. The lower four panels show the corresponding full spectral scans through the cores of the same spectral lines, and also that of the Ca~II 854.2~nm lines at four spatial locations indicated with colours that correspond to those of the spectral scans. The black dashed line corresponds to the line profile averaged over the entire FOV. The data shown are from \citet{2025A&A...696A...3L}. }
\label{fig:AR_spectra}
\end{figure*}

In Figs.  \ref{fig:AR+plage}--\ref{fig:AR_spectra} we show examples of images and processed data from CRISP and CHROMIS. Figure \ref{fig:AR+plage} illustrates the quality of images from the most frequently employed spectral lines used by these instruments: the Ca~II K line and H$\beta$ with CHROMIS, and the Fe~I 617.3~nm line (used to infer photospheric magnetic fields and dynamics)  and the Ca~II 854.2~nm line (used to infer chromospheric dynamics and magnetic fields) with CRISP. Figure \ref{fig:LOS_magnetic_field} shows an image and a LOS chromospheric magnetic field obtained from the Ca~II 854.2~nm line (top), an image in the continuum adjacent to the Fe~I 617.3~nm line, and a map of the LOS photospheric magnetic field obtained from the same line. Figure \ref{fig:AR_spectra} shows images selected from spectral lines indicated in the Figure, and the corresponding full spectral scans through these lines at four selected positions, indicated with coloured plus signs in the upper panels. The data presented in these figures define the present state of the art in narrowband imaging and spectropolarimetry with ground based solar telescopes. In the following we describe how CRISP and CHROMIS are designed and how similar concepts can be carried over to the future EST.

\begin{table*}[tbh]
 \caption{\label{table_exp1}Input parameters for the seven dual FPI systems discussed.}
  \centering
  \small
  \begin {tabular} {lccccccc}
    \hline
    \mathstrut
    Parameter / FPI system & CRISP & CRISPm & CRISP2 & CHROMIS & EST-B & EST-V & EST-R \\ 
    \hline
    Telescope & SST & SST & SST & SST & EST & EST & EST \\
     FPI system status  & Obsolete & Operational & In progress & Operational & Proposal & Proposal & Proposal\\
     Telescope diameter (m) & 0.98 & 0.98 & 0.98 & 0.98  & 4.2 & 4.2 & 4.2 \\
     Input focal ratio & 46 & 46 & 46 & 46 & 50 & 50 & 50 \\    
     FOV diameter (arc min) & Max & Max & Max & Max & 1.0 & 1.0 & 1.0 \\ 
     Etalon clear aperture (mm) & 75 & 75 & 98 & 75 & Min & Min & Min \\ 
     Camera pixel size ($\mu$m) & 16 & 6.5 & 6.5 & 5.48 & 5--12 & 6--12 & 6--12\\     
     Wavelength range (nm) & 500--860 & 500--860 & 520--860 & 390--500 & 380--500 & 500--680 & 680--1000\\
     Reference wavelength(s) (nm) & 630 & 630 & 630 & 390 & 380,396 & 500,630 & 680,854 \\       
     Required  rms cavity error (nm) & 2 & 2 &2 &2 & 2 & 2 & 2 \\
     Target Strehl intensity (\%) & 90 & 90 & 90 & 90 & 90 & 90 & 90 \\
     Spectral FWHM at ref. wavel.  (pm) & 6 & 6 & 7 & 8 & 7.9 & 6.3 & 11\\
     Peak transmission (\%) & 90 & 90 & 90 & 90 & 90 & 90 & 90 \\
     Parasitic light level (\%) & 1.0 & 1.0 & 1.0 & 1.0 & 1.0 & 1.0 & 1.0 \\
     Prefilter type & 2-cavity & 2-cavity & 2-cavity & 2-cavity & 2-cavity & 2-cavity & 2-cavity \\
 \hline
  \end{tabular}
  \tablefoot{
  CRISP and CRISPm are identical except for their camera lenses. The three designs of FPI systems for EST have been proposed by \citep{2025arXiv250521053S}, based on requirements decided by the EST Science Advisory Group, but have not been reviewed by the EST project. The first reference wavelength for EST FPI systems refer to the requirement on Strehl, the second to the requirement on spectral resolution. The Strehl intensities required refer to a combination of apodisation effects and optical design limitations.}
\end{table*}

\begin{table}[tbh]
\caption{\label{table_exp_CRISP2}Comparison of two similar dual FPI systems}
  \centering
  \small
  \begin {tabular} {lccccccc}
    \hline
    \mathstrut
    \hspace{3.5cm}FPI system: & Conventional & Improved  \\
        \hline
    {\bf Input parameters:}\\    
    Cavity separation $d_1 (mm)$ & 0.787 & 0.787 \\
    Cavity separation $d_2 (mm)$ & 0.300 & 0.300 \\
    Reflectivity FP1 (\%) & 93 & 93 \\
    Reflectivity FP2 (\%) & 93 & 86 \\
    Prefilter FWHM (nm) & 0.43 & 0.43 \\
    Assumed rms cavity error (nm) & 2 & 2 \\
    \hline
    {\bf Calculated performance parameters:}\\
    Strehl from pupil apod. at 500 nm (\%) & 90.9 & 94.5 \\
    Transmission at peak $\lambda$ (\%) & 81.4 & 93.5 \\
    rms Integrated transmission (\%) & 15.4 & 6.4\\
    Broadened FWHM at 630 nm (pm) & 6.5 & 6.7 \\
    rms FWHM (pm) & 0.44 & 0.064 \\
    rms wavelength shift (pm) & 1.5 & 1.6 \\
    rms asymmetry (pm) & 0.87 & 0.55 \\
    Parasitic light (\%) & 0.25 & 0.79\\
    rms Parasitic light (\%)& 0.055 & 0.062 \\
    \hline
  \end{tabular}
  \tablefoot{ 
  The first column corresponds to an FPI system with a low-resolution etalon that has high reflectivity and the second column to a system with a low reflectivity low-resolution etalon, all other input parameters are identical for the two systems. The system with low reflectivity for the low resolution etalon demonstrates superior performance compared to the other system. Calculations are made with input parameters for CRISP2, which are very close to those of the EST-V FPI system.}
\end{table}

\begin{table*}[tbh]
\caption{\label{table_exp2}Calculated parameters for seven dual FPI systems.}
  \centering
  \small
  \begin {tabular} {lccccccc}
    \hline
    \mathstrut
    Parameter / FPI system & CRISP & CRISPm & CRISP2 & CHROMIS & EST-B & EST-V & EST-R \\
     \hline
    Total length of FPI system (m) & 1.437 & 1.316 & 1.824 & 1.579 & 4.4--4.7& 4.5--4.7 & 4.3--4.5\\ 
    Etalon clear aperture (mm) & (75) & (75) & (98) & (75) & 135 & 180 & 180 \\
    FOV diameter (arc min) & 1.5 & 1.5 & 2.4 & 2.0 & (1.0) & (1.0) & (1.0) \\
    Focal ratio at etalons & 165 & 165 & 140 & 120 & 110 & 147 & 147\\
    Output focal ratio & 55 & 23 & 20 & 31 & 25--60 & 25--46 & 19--35 \\
    Cavity ratio $d_2/d_1$ & 0.381 & 0.381& 0.381& 0.381& 0.381& 0.381& 0.381 \\
    FP1 cavity gap $d_1 (mm)$ & 0.787 & 0.787 & 0.787 & 0.358 & 0.377 & 0.787 & 0.787   \\
    FP2 cavity gap $d_2 (mm)$ & 0.300 & 0.300& 0.300 & 0.137 & 0.144 & 0.300& 0.300 \\
    FP1 Reflectivity (\%) & 94* & 94* & 93* & 90* & 90 & 93 & 93 \\
    FP2 Reflectivity (\%) & 86* & 86* & 86* & 80* & 80 & 86 & 86 \\
    \hline
    Parameters at ref. wavelength(s)  \\
    \hline
    Strehl from optical design  (\%) & 97 & 96 & 93 & 92 & 93 & 97 & 97 \\
    Strehl from pupil apodization  (\%) & 96 & 96 & 96 & 97 & 96 & 95 & 97 \\
    Unbroadened FWHM (pm) & 4.9 & 4.9 & 5.7 & 6.8 & 6.8 & 5.7 & 10.4  \\ 
    Broadened FWHM (pm) & 5.7 & 5.7 & 6.9 & 7.9 & 7.9 & 6.7 & 11.4 \\
    rms FWHM (pm) & 0.043 & 0.043 & 0.064 & 0.083 & 0.083 & 0.064 & 0.081 \\
    rms wavelength shift (pm) & 1.6 & 1.6 & 1.6 & 2.1 & 2.1 & 1.6 & 2.1 \\
    Peak FPI transmission (\%) & 93.5 & 93.5 & 93.5 & 92.5 & 92.5 & 93.5 & 96.3\\
    rms Integrated transmission (\%) & 6.6 & 6.6 & 6.4 & 7.5 & 7.5 & 6.4 & 3.8\\    
    Parasitic light (\%) & 0.75 & 0.75 & 0.79 & 1.0 & 1.0 & 0.79 & 0.77\\
    rms Parasitic light (\%) & 0.062 & 0.062 & 0.062 & 0.096 & 0.096 & 0.062 & 0.033 \\
    Prefilter FWHM (nm)  & 0.43 & 0.43 & 0.43 & 0.28 & 0.28 & 0.43 & 0.80\\
  \hline
  \end{tabular}
  \tablefoot{
  The calculated parameters are based on the input parameters given in Table \ref{table_exp1}, optimisation of their optical design, numerical simulations, and optimisation of FPI parameters. The Strehl intensities based on apodisation effects and optical design limitations are calculated separately. * Reflectivities given are design values, actual values differ from these.}
\end{table*}

\section{Conceptual design of seven solar FPI systems} \label{Conceptual_designs}
In the following, we describe the steps taken to design and construct four dual FPI systems: CRISP, in its original version and a modified version with a new camera lens that we refer to as CRISPm, CRISP2, and CHROMIS. We propose to adopt the same strategy to the design of three similar FPI systems for EST, dubbed EST-B, EST-V and EST-R. The first step is to define the input optical system and various target values for the performance of the systems. We give such input parameters for all seven systems in Table \ref{table_exp1}. It should be remarked that some of the requirements set for EST by its Science Advisory Group (SAG) differ slightly from those given in the Table. This is because the required spectral FWHM for CRISP, CRISPm, CRISP2 and EST-V are almost identical, and in order to facilitate a comparison between the designs of these systems, we have assumed the same cavity separations for these systems but maintained the actual and proposed focal ratios. The same holds for CHROMIS and EST-B, for which we assume the same cavity separations but slightly different focal ratios. Furthermore, the spectral resolution of CRISP, CRISPm, and CRISP2 is almost identical to that of EST-R at a wavelength of 854 nm, so we merge also these requirements. Taken together, this means that the proposed systems for EST are not exactly those required by SAG but at the same time, we expect other differences compared to the present proposal as well, if and when these systems are built. The steps used to design the FPI systems are the following:
\begin{itemize}
\item Establish input parameters and other constraints, see Table \ref{table_exp1}. For the EST FPI systems, we have optimised optical designs for two pixel sizes, either 5~$\mu m$ and 12~$\mu$m, or 6.5~$\mu$m and 12~$\mu$m, constituting our best guess of the range of future pixel sizes of suitable cameras.  
\item Given a target Strehl of  95\%, by accounting only for the image quality degradation by pupil apodization and pupil phase errors at the required spectral resolution (see next item), establish the needed minimum F-ratio of the telecentric system at the location of the etalons. This and the required FOV defines the needed clear aperture of the etalons and thus also the size of the optics. For SST, the etalon clear aperture diameters were given, and we instead maximised the FOV diameter with that clear aperture. For these calculations to be realistic, the low resolution etalon is tilted, which is needed to eliminate ghost images.
\item  In parallell with the previous task, define the reflectivities and cavity separations that deliver the required spectral FWHM, given the maximum rms cavity errors agreed with the manufacturer of the etalons. This involves identifying and choosing a cavity ratio that minimises parasitic light (spectral stray light from side lobes), and fixing the FWHM of the prefilter.
\item  Given the dimensions of the etalons and optics, and the focal ratio of the beam at the etalons, design the optical system of the narrowband system to be as compact as possible, while allowing a Strehl limited by the optical design of 95\%. Include wideband beam splitter on the input side and polarising beam splitter on the output side, and the polarisation modulator if a design exists.
\item  Check that the design is short enough to fit within the available space. For CRISP, the initial constraint was an overall length of 1.5 m, for EST the absolute maximum length is 6~m including cameras, and we are targeting a significantly shorter length. Also, check for other constraints on space for etalons and for a polarisation modulator (see Sect. \ref{EST_particularities}).
\item  Design the wideband system to deliver the same image scale and the same image quality as the narrowband system. The default solution is to use duplicates of the first lens and the last (camera) lens of the narrowband system as first and second lenses of the wideband system, but for EST we found that modifications of the camera lens design for the wideband system is preferable.
\item  Carry through tolerance analysis for the manufacture and mounting of the lenses, establish with a potential contractor for the lenses that these tolerances are realistic.
\item  Proceed with the overall mechanical design of the narrowband and wideband systems
\end{itemize}

\subsection{Simulations and optimization of FPI parameters} \label{simulations}
\subsubsection{Overview of simulations}
The software used for these simulations is by and large the same as used for the calculations presented by \citet{2006A&A...447.1111S}. The code consists of essentially two parts that are independent of each other. One part calculates the PSF integrated over the passband by taking into account the complex transmission of the pupil from multiple internal reflections in the cavity, leading to accumulated phase errors and pupil apodization (from the wavelength shift of the peak transmission of tilted rays) at the given focal ratio at the location of the etalons. These calculations take into account the tilt of the low-resolution etalon, which needs to be $1/(2 F\#)$~radians, where F\# is the focal ratio, to remove ghost images with a pupil stop located at the pupil formed by the first lens behind the two FPIs. The monochromatic PSF is calculated for one wavelength at the time within the passband of each FPI. By integrating the monochromatic PSFs over wavelength, the integrated PSF is obtained, from which we obtain a Strehl by comparing to an identically integrated PSF but obtained with a perfectly collimated beam (infinite focal ratio). To allow for focus compensation, a phase term with variable amplitude $c_4$ is added to the accumulated phase errors of the FPI system:
\begin{equation} 
  \phi_4=c_4 \sqrt{3} (2r^2-1),
\end{equation}
in units of radians and where $r$ is the distance from the center of the pupil, and the Strehl is calculated for a number of values of $c_4$. From this set of Strehl values, the peak value is found by quadratic interpolation. 

A major result from our calculations was that focus compensation improves the Strehl of FPIs in telecentric mounting in a dramatic way, such that the phase errors are actually of minor importance in comparison with the apodisation effects \citep{2006A&A...447.1111S}. Our calculations of the Strehl made without focus compensation were compared to those of \citet{2000A&AS..146..499V}, and show discrepancies of 2\% or less. The calculations of Strehl values after optimisation by focus compensation found by \citet{2006A&A...447.1111S} were much later confirmed with a different method of computation by \citet{2010A&A...515A..85R} and found to agree to within 1--2\%. We believe that this corresponds to the level of accuracy of our Strehl calculations. Since these calculations were made, several minor improvements of the code were made but only one significant bug was found. The bug was that when tilting the low-resolution etalon, the associated wavelength shift of the transmission peak of that etalon was not compensated for by retuning that etalon in wavelength. This caused the low-resolution etalon to be slightly mistuned relative to the high-resolution etalon. This bug had only a minor effect on the results with tilted low-resolution etalons but no effect at all on the results referred to above.

The above described part of the code delivers a Strehl and a spectral FWHM of the transmission profile, which is broadened by the effects of the tilted rays at the pupil. During optimisation of the system, we check whether the obtained broadened FWHM differs from its target value. If so, the cavity separations are adjusted and the Strehl calculations repeated. Similarly, the focal ratio is adjusted to provide a Strehl that is close to the target value of 95\%. This process converges quickly.

The second part of the code takes into account the effects of cavity errors, assumed to have a Gaussian distribution with an rms variation of 2 nm, on the performance of the FPI system. This corresponds to the cavity errors actually measured for both etalons of CRISP (and CRISPm), and what is expected from the etalons for CRISP2. The effects of these cavity errors, with a telecentric mounting of the etalons, are quasi-random wavelength shifts of their transmission peaks over the FOV, which causes localised and variable mistuning between the transmission peaks of the high- and low-resolution etalons. Even with relatively small cavity errors, this can seriously degrade performance of the FPI system, because it causes an overall drop in the transmission of the system and also leads to (widely) varying properties of the combined spectral transmission profile over the FOV. This in turn can increase the sensitivity of the transmission profiles to small tuning errors, which tends to disrupt the robustness of the system and also may have adverse consequences for the interpretation of the science data. A second major result of our previous simulations is that these destructive effects of cavity errors on the performance of dual FPI systems to a large extent, and at relatively small cost, can be mitigated simply by lowering the reflectivity of the low-resolution etalon \citep{2006A&A...447.1111S}. This widens the low-resolution transmission profile to compensate for the relative wavelength shifts of the peaks of the two FPIs. The price paid is enhanced spectral straylight from the side lobes of the transmission profile, but that can be balanced by adjusting the FWHM of the prefilter. 

The simple but important trick of lowering the reflectivity of the low-resolution etalon has a profound impact on the robustness and fidelity of the performance of dual FPI systems, which we illustrate in Table \ref{table_exp_CRISP2} with parameters calculated for CRISP2. These parameters are also very close to the parameters of CRISP (thus also CRISPm), and what is proposed for EST-V. Here, we compare calculated properties of two systems, which are identical in all respects except for the reflectivity of the low-resolution etalon. In the first column, this reflectivity is 93\% and in the second column it is 86\%, which is also the reflectivity of the high-resolution etalon. The Strehl improves from 91\% to 95\%. The "Transmission at peak wavelength" in this Table corresponds to the transmission of the system at the wavelength of the peak transmission of the high-resolution etalon, and is increased from 81\% to 94\%. We consider this to be a "spectral fidelity number", of an importance that is comparable to that of the Strehl value, and in our designs we target a value of at least 90\%. The integrated transmission corresponds to the integral of the integrated transmission profile over wavelength and would correspond to the intensity measured at a continuum wavelength. Its rms variation over the FOV is reduced by a factor 2.4 with the reduced reflectivity of the low-resolution etalon, thus predicting a reasonably smooth variation of the baseline intensity over the FOV. The rms variation of the FWHM of the spectral transmission profile shows a major improvement by nearly a factor 7 and is only 1\% of its average value after lowering the low-resolution FPI reflectivity. We consider this to be the most important signature of a robust dual FPI system. Also, the variability of the asymmetry of the transmission profile is reduced. These improvements are at the (small) price of an increased parasitic light level (spectral straylight from side lobes) from 0.25\% to 0.79\%. At the same time, the relative variation of this parasitic light over the FOV is reduced by nearly a factor 3, which suggests the possibility of calibration and removal of this effect from the data in postprocessing, if necessary.

\subsubsection{Optimized parameters for seven solar FPI systems}
Based on the input parameters defined in Table \ref{table_exp1}, and with additional requirements of a target Strehl calculated from FPI effects (not including limitations from optical design) close to 95\%, and a required transmission at the peak wavelength of at least 90\% (as discussed in Sect. \ref{simulations}), we obtain the cavity separations and reflectivities of seven dual FPI systems, as summarised in Table \ref{table_exp2}. We also obtain the required FWHM of their 2-cavity prefilters, needed to reach the 1\% parasitic light level. This Table also provides predicted parameters of several additional parameters that quantify and establish the expected high level of robustness and fidelity of these FPI systems, as explained in Sect. \ref{simulations}.

We emphasise, that the cavity separations and reflectivities of the dual etalon systems proposed for EST are very similar to those of CRISP and CHROMIS, which have been in operation at SST for many years. For EST-V and EST-R, we propose 93\% and  86\% reflectivities for the high- and low-resolution etalons, but for EST-B, we propose 90\% and 80\% reflectivities to account for the stronger effects of cavity errors at wavelengths below 400~nm. We have assumed 2~nm rms cavity errors for the EST etalons as for those of SST, which is demanding. In comparison, the single etalon version of the FPI system of DKIST, the Visible Tunable Filter \citep[VTF; ][]{2012SPIE.8446E..77K, 2014SPIE.9147E..0ES} is reported to show 3~nm rms cavity errors over its 250~mm clear aperture \citep{2024SPIE13096E..17H}, suggesting that we should expect large etalons to have larger cavity errors than small etalons. The clear apertures for the EST etalons are significantly smaller than those of VTF: EST-V and EST-R need 180 mm clear apertures to reach the same spectral resolution and FOV diameter as VTF, and EST-B needs only 135 mm etalon, thanks to its lower spectral resolution, which is almost identical to that of CHROMIS.

Summarising, it can be concluded that the feasibility of making the required etalons for EST is supported by the existence of dual etalon systems at SST (CRISP and CHROMIS) with almost identical cavity separations and reflectivities, but about half their diameters. The availability of an even larger high-quality etalon built for DKIST confirms this conclusion. The required cavity errors of the etalons for EST appear realistic and have been verified on CRISP and CHROMIS while in operation at SST.
 
\subsection{Telecentric FPI system}
\begin{figure*}[htbp]
\center
\scalebox{1.05} [2.4]{\includegraphics[angle=0, width=0.7\linewidth,clip]{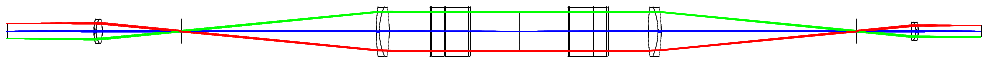}}
\scalebox{1.05} [2.05] {\includegraphics[angle=0, width=0.7\linewidth,clip]{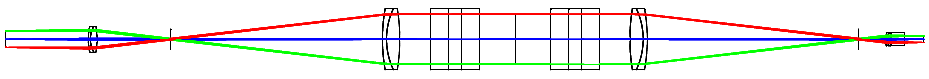}}
\scalebox{1.05} [1.07] {\includegraphics[angle=0, width=0.7\linewidth,clip]{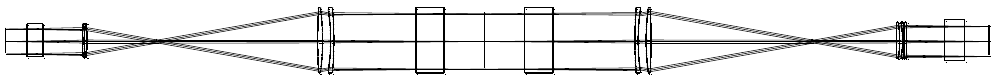}}
\scalebox{1.00} [1.07] {\includegraphics[angle=0, width=0.75\linewidth,clip]{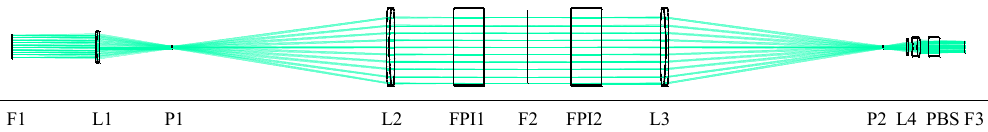}}
 \caption{
Layout of four of the FPI systems discussed in this paper, here shown scaled individually to highlight similarities in their optical designs. The systems shown are from top to bottom: CHROMIS (overall length 1.579~m, FPI clear aperture diameter 75 mm), CRISP2 (1.824~m, 98 mm), EST-B (4.4--4.7~m, 135~mm), and EST-V (4.5--4.7~m, 180~mm). Symbols used: F1-F3 are focal planes, L1-L4 lenses, P1-P2 pupil planes, PBS polarising beam splitter. The vertical scale of all panels has been expanded 2$\times$ for clarity. For further details about the systems, see Tables \ref{table_exp1} and \ref{table_exp2}.}
\label{fig:FPI_layouts}
\end{figure*}

The optical designs of CRISP, CRISP2 and CHROMIS, as well as the three FPI systems proposed for EST all use a combination of two telecentric imaging systems to transfer the image from a primary focal plane through the FPI system and onto the cameras. The particular points of interest in these designs are indicated in Fig. \ref{fig:FPI_layouts} and are: the primary focal plane F1, the first lens L1, the first pupil plane P1, the second lens L2, the secondary focal plane F2, close to which the etalons are located, the third lens L3, the second pupil plane P2, the fourth lens L4, often referred to as the camera lens, and the camera focal plane F3.

In our FPI systems, L2 and L3 have identical designs. There are thus three degrees of freedom in the design and these are the focal lengths of L1, L2/L3, and L4. These are given by the needed F-ratio at the etalons (Sect. \ref{simulations}), the image scale at the camera focal plane, and the total length of the system from F1 to F3. In our designs, we strive to make the system as compact as possible while maintaining a Strehl, limited by system aberrations and field curvature, close to 95\%. We also try to make the system robust in three respects: one is by paying attention to manufacturing and alignment tolerances, the other allowing change of image scale by only replacing the camera lens (L4), without degrading image quality - thus is in order to enable simple upgrades of cameras in the future. The third aspect is to ensure stability of the reimaging system with respect to a reasonable range of temperatures.

The FPI system needs a beam splitter, which can be located either before or after L1, to deflect 5-10\% of the light to the wideband system, and a polarising beam splitter between L4 and F3. The dimensions of these are set by the input and output focal ratios and the diameter of the FOV, and both are included in the optical design. There is also a polarisation modulator that is placed either before L1 or between L4 and the polarising beam splitter. For the EST FPI systems we do not yet have an optical design for the modulator, so that is not part of the FPI optical design. Our previous experience is that the modulator only shifts the focal plane without any degradation of image quality, so that is an addition that can be made later.

Both pupil locations have slightly oversized pupil stops that are of critical importance for eliminating ghost images, stray light and spurious interference fringes. At P1, a concern is reflections involving the prefilter, on the output side the low-resolution etalon is tilted just enough to allow a pupil stop at P2 to completely eliminate ghost images from inter-etalon reflections.
  
During the design, we pay attention to the following aspects, in addition to those that are common practice in lens design, such as minimising angles of incidence to relax assembly tolerances:
\begin{itemize}
\item Telecentricity of the beam at the FPIs to ensure their proper functioning.
\item Telecentricity of the output beam to minimise image scale changes when refocusing.
\item Parallelism of light between large and small achromats to minimise image scale changes when refocusing.
\item The limited availability of optical glasses with high transmission for large lenses, in particular at wavelengths shorter than 400~nm (CHROMIS and EST-B). If the choice is to cement large lenses, then matching of their coefficients of thermal expansion is a challenging problem.
\end{itemize}

A source of straylight that should also be taken into account is the light reflected by the camera sensor, and back to the second (low-resolution) etalon. With an appropriate tilt of the camera, this light will not pass through the second pupil stop. With SST, the focal plane is actually tilted because of the off-axis arrangement of its Schupmann system, so tilting the cameras does not degrade image quality - on the contrary. But with telescopes such as EST, it may be worthwhile to arrange its secondary optical system (POP) to deliver a slightly tilted focal plane. 

\subsubsection{Particular challenges related to EST}  \label{EST_particularities}
Designing FPI systems for EST faces particular challenges, in part because of the large diameter of EST (4.2~m), and in part because of its secondary optical system, referred to as the Pier Optical Path  \citep[POP;][]{2022A&A...666A..21Q}. This is the optical system that transfers the image from the telescope to the Coudé rooms, which corresponds to a distance of about 30~m. The challenges of the design of this system come from this large distance in combination with the relatively fast output beam of the transfer optics and calibration assembly (TOCA) that provides the input to POP. The solution was to design POP with a large triplet lens, a dichroic beam splitter that divides the input into two beams with a dividing wavelength of 680~nm, plus a large doublet lens for each of the two beams. Practical limitations of this solution forced the output beams of POP to be slow, with focal ratios of about 77 for the shorter wavelength beam and 60 for the other beam. 

For the FPI systems, a slow POP beam causes problems with large diameters of the pre-filters, which are located close to the output focal plane of POP. This and problems related to the manufacture of the large lenses of POP triggered a revision of its optical design, which we support strongly. By adding a singlet field lens in the interior of POP, the lens diameters could be strongly reduced and a focal ratio of 50 could be reached for both beams of POP. Our design of FPI systems uses a redesign of POP by Álvaro Pérez García, which is telecentric or nearly telecentric at all wavelengths. This allows the use of prefilters with diameters of 70~mm for a FOV diameter of 1 arc min., which by leading manufacturers of such filters is considered to be well within their present capabilities, and expected to not constitute a major cost driver.

A major challenge for the FPI systems of EST is the focus curve of POP, which needs fast focus compensation in order to not reduce the overall duty cycle and efficiency of observations made by alternating between two or more spectral lines with any particular FPI system. The time at the disposal for such refocusing, without loss of efficiency, corresponds to the time needed for changing prefilter, which is necessary when tuning to a new spectral line, and may take a fraction of a second. There are several possibilities for refocusing, including refocusing of POP, which is difficult because of the fact that both lenses of POP act as vacuum windows, but also because of the large mass involved in moving the lenses with their mounting. Another possibility is to move the entire mechanical construction holding the FPI system with its wideband system and cameras, but that involves moving a weight of probably several hundred kilos a distance of about 10~mm in a fraction of a second. A third possibility is to use the first lens of the FPI system, but this moves the pupil image and also has a very negative impact on the telecentricity of the system at the location of the etalons, so that option does not appear viable.

The fourth possibility is to move either the camera or the camera lens for refocusing. Simulations show that moving the camera in principle could work even better than moving the camera lens, but we nevertheless advice against that solution for two reasons: the first is that a narrow- and wide-band combined FPI system will contain four cameras and thus needs four motorised translation stages per FPI system. The other objection is that moving cameras with connected cables for data transfer and perhaps cooling appears more cumbersome than moving a small lens. We therefore recommend to move the camera lens to compensate for the focus curve of POP. This is a good solution because:
\begin{itemize}
  \item This lens is small, thus involves only a small mass to move.
  \item The needed movement is small (but scales as the square of the output focal ratio).
  \item Rapid and highly repeatable linear translation stages are commodity hardware available from several manufacturers of optomechanics.
  \item The reimaging system is telecentric, such that refocusing involves only a small change in image scale.
\end{itemize}

Our concern with the focusing mechanism of the FPI systems is not the focus curve of POP, but to limit the overall needed focus range by minimising focus errors on the input side, such as those from thermal drifts. This is discussed in Sect. \ref{EST-B_telecentric_design}.

\subsubsection{Wideband system}
The proper functioning of the wideband system is of crucial importance for the processing of narrow-band data obtained from wavelength scans across spectral lines, with or without polarimetry (Sects. \ref{wideband_images} and \ref{image_restoration}). The optical quality of the wideband image needs to be at least as good as that of its corresponding narrow-band system. At present, our recommendation is also that the image scale of the wide-band system should closely match that of the narrow-band system, because this is a requirement of the existing CRISP and CHROMIS data processing pipeline and MOMFBD code \citep{2002SPIE.4792..146L, 2005SoPh..228..191V, 2015A&A...573A..40D, 2021A&A...653A..68L}. However, there is no profound reason why this should be needed, as long as subfielding and co-alignment of images from the two systems allow aberrations in seeing degraded image pairs to be discerned and compensated for (Löfdahl, private communication). Nonetheless, rewriting the software to manage different image scales of the two systems is likely a substantial effort. Before to undertake such an efforts, there are two question to be answered through simulations: one is what percentage of discrepancy in image scales the MOMFBD code can handle without significant negative impact on the restored images. The other is whether in fact MOMFBD processing can be made in two steps instead of one by first processing the wideband images and then using the result from that processing to process the narrowband images (Löfdahl, private communication). If so, there is no need to rewrite the MOMFBD code to handle unequal image scales.  In this context, it is worth noting that both \citet[][applied to spectral reconstructions]{2017A&A...608A..76V} and \citet[][applied to MiHi data]{2022A&A...668A.151V} process data first from their wideband camera and then from the "narrowband" (spectral or MiHi data) camera and this is reported to work very well. This suggests that a similar separate processing of wideband and narrowband FPI data should also work well. 

Until the above questions have been answered, our approach is to design the wideband with an image scale that follows that of the narrowband system as closely as possible unless that leads to loss of image quality or a complex design of the wideband camera lens. For CRISP, CRISPm, CRISP2 and CHROMIS we have successfully used duplicates of L1 and L4 to construct the corresponding wideband systems but for the EST FPI systems, minor modifications of the design of the wideband camera lenses were needed to fully comply with the requirements. 

\subsection{Optical design of seven telecentric FPI systems}

Table \ref{table_exp2} summarises the main results of the optimisations of the seven FPI systems discussed in this papers. This Table provides the estimated minimum Strehl values obtained separately from limitations set by pupil apodisation and phase errors associated with multiple reflections in the etalon cavities, and from the optical design of the telecentric re-imaging system. This Table indicates that, as regards the optical design, the most demanding systems are CRISP2, CHROMIS, and EST-B, which achieve Strehl values that are 2--3\% below the target 95\%. The challenge of CRISP2 is its large FOV diameter, 2.3~arc min combined with its overall length, which is constrained by the space available in the SST optics lab. The challenge of CHROMIS and EST-B is  the limited availability of large blanks of high-quality optical glass with high transparency at 380~nm.

In the following, we discuss the details of the optical designs of only three systems: first, CHROMIS \citep{2017psio.confE..85S}, which primarily aims at observing the solar chromosphere in the Ca~II H and K lines, and in H$\beta$, and is now (May 2025) being upgraded with polarimetry. CRISP2, which is in the final stages of construction and intended to replace CRISPm at SST and play a unique role as the only imaging spectropolarimeter that combines high spatial resolution with a large FOV. With the design of EST-B, which is part of a proposal of three FPI systems for EST \citet{2025arXiv250521053S}, we explore some of  the challenges associated with the focus curve of EST, through its secondary re-imaging system POP. The design of remaining FPI systems for EST, EST-V and EST-R, are presented in a separate document \citep{2025arXiv250521053S}.

\subsubsection{CHROMIS telecentric optical design}
 CHROMIS \citep{2017psio.confE..85S} is the second dual FPI system installed at SST on La Palma, and its primary purpose is to observe the chromospheric Ca~II H and K lines at 396.8~nm and 393.4~nm, and H$\beta$ at 486.1nm. CHROMIS was built to provide polarimetric capabilities but the polarimeter was only recently (May 2025) installed. Because of the very low count rates in the core of the Ca~II lines, the spectral resolution aimed for was 50,000, or about 8~pm, which is sufficient to distinguish the blue and red emission peaks from the darker core in the quiet Sun, thus providing some information about LOS velocity gradients in the chromosphere. There is strong evidence to suggest that the reflectivities of the etalons are lower than indicated by direct measurements, and that therefore the spectral resolution is below its target. However, given the low light levels even at that lower spectral resolution, the decision was to not recoat the etalons. 

\begin{table}[tbh]
\caption{\label{table_CHROMIS}Prescriptions of the narrowband lens system of CHROMIS.}
  \centering
  \small
  \begin{tabular}{lrrrc}
    \hline
    \mathstrut
Lens & Radius & Thickness  & Lens dia & Glass \\
 & (mm) & (mm) & (mm) & \\
\hline
L1& 75.80 &3.4 &40 & S-BSM10 \\ 
L1 & 37.80 & 10.2 & 40 & S-FPL53 \\
L1 & -95.81 & 129.3 & 40  \\
L2/L3 & 292.26 & 5.0 & 80 & PBM18Y \\
L2/L3 & 102.74 & 16.0 & 80 & BSL7Y \\
L2/L3 &-285.24 & 66.7 & 80 \\
L4 & 85.95 & 8.0 & 30 & S-FPL51Y \\
L4 & -28.32 & 3.5 & 30 & PBL6Y \\
L4 & -74.38 & 91.6 & 30 \\ 
 \hline
  \end{tabular}
  \tablefoot{Of the four lenses of the narrowband system of CHROMIS, L2 and L3 are identical. All lenses are cemented doublets. Duplicates of L1 and L4 are used to build the corresponding wideband system of CHROMIS,}
\end{table}

 The low spectral resolution of CHROMIS allows an F-ratio of 120 at the etalons, which delivers a larger FOV (about 2 arc min diameter) than CRISP with 75~mm etalons, but smaller than CRISP2 (about 2.4 arc min diameter) with its 98 mm etalons. The design of CHROMIS  allowed utilisation of its the entire FOV but the initial cameras sampled only 75$\times$47 arc sec. The cameras installed in May 2025 sample 80$\times$80 arc sec and are described in Sect. \ref{CHROMIS_cameras}.
 
The final optical design of CHROMIS uses actual dimensions of the FPIs  and highly transmitting glass for the four lenses of the system (Table \ref{table_CHROMIS}). The transmission losses caused by absorption in the glass is estimated to be 2.5\% at 390~nm and 1\% AT 490~nm. With estimated 0.3\% reflections after anti reflection coatings, the 12 air to glass surfaces (including the FPIs) are expected to lead to 3.5\% additional loss. The lowest Strehl over the FOV in the 390-490~nm wavelength range is 93\%. A detailed tolerance analysis combined with simulations suggest that manufacturing and alignment errors could lower this Strehl by about 5\%. The design also involves a detailed study of possible ghost images from the various flat surfaces of the FPIs, and the external surfaces of the etalons are wedged by an angle that is sufficiently large to prevent these ghost images to pass through the pupil stops.

The blue beam at SST receives light from a dichroic beam splitter that reflects light shortward of about 500~nm. A specially designed double beam splitter feeds 90--95\% of this light to the FPI system, while the reminder is divided between the wideband system and a correlation tracker camera that controls the SST tip-tilt mirror \citep{2024A&A...685A..32S}.

CHROMIS is, to our knowledge, the first FPI based narrowband filter system capable of observing the Sun at wavelengths below 400~nm. In Figs. \ref{fig:AR+plage} and \ref{fig:AR_spectra}, we illustrate the achievable image quality with images recorded in the core of the Ca~II K line, the nearby continuum at 400.0nm, and H$\beta$.

\subsubsection{CRISP2 telecentric optical design}
CRISP2 represents the next second of imaging spectropolarimeters for SST. The layout is shown in Fig. \ref{fig:FPI_layouts}. It is designed to cover a niche in parameter space that the generation of 4--meter solar telescopes, such as DKIST and the future EST, will have difficulties to fill: that of high-spatial resolution observations over a large FOV. The design of CRISP2 therefore is designed with priority to cover as large FOV as possible with 100~mm clear aperture etalons. Given the spectral resolution required, the decision was to design the system with a focal ratio of 140 at the etalons, which allows a FOV diameter of  nearly 2.4~arc min. Furthermore, the science priority is observations at relatively long wavelengths, such as obtaining spectropolarimetric data in the 617.3~nm, 630.2~nm and 854.2~nm lines, with additional observations of chromospheric dynamics in H$\alpha$, which permitted a small reduction of the Strehl in the outermost parts of the FOV at the shortest wavelengths accessible with CRISP2, see Table \ref{table_CRISP2_strehl}.

 \begin{table}[h]
 \caption{\label{table_CRISP2_strehl}CRISP2 Strehl values.}
  \centering
  \small
     \setlength{\tabcolsep}{4.5pt}
  \begin{tabular}{rrccccccccc}
        \hline
    \mathstrut
    $\lambda$ &(nm)& 530 & 550 & 600 & 650 & 700 & 750 & 800 & 850 \\
    x & y\\
    \hline
0 & 0  & 0.99 & 0.98 & 0.97 & 0.95 & 0.95 & 0.94 & 0.95 & 0.96 \\
0 & 60"  & 0.96 & 0.96 & 0.97 & 0.98 & 0.98 & 0.98 & 0.98 & 0.98 \\
0 & -60"  & 0.96 & 0.96 & 0.97 & 0.98 & 0.98 & 0.99 & 0.98 & 0.98 \\
60" & 0 & 0.98 & 0.97 & 0.97 & 0.97 & 0.97 & 0.96 & 0.96 & 0.95 \\
-60" & 0  & 0.94 & 0.95 & 0.98 & 0.99 & 0.99 & 1.00 & 0.99 & 0.99 \\
0 & 70"  & 0.92 & 0.93 & 0.94 & 0.95 & 0.96 & 0.97 & 0.97 & 0.96 \\
0 & -70"  & 0.93 & 0.93 & 0.94 & 0.96 & 0.97 & 0.97 & 0.97 & 0.96 \\
70" & 0  & 0.97 & 0.97 & 0.96 & 0.96 & 0.96 & 0.95 & 0.95 & 0.94 \\
-70" & 0  &0.88 & 0.91 & 0.94 & 0.97 & 0.98 & 0.99 & 0.99 & 0.98 \\
 \hline
  \end{tabular}
  \tablefoot{CRISP2 Strehl values at different x and y offsets in arc sec from the center of the FOV, and wavelengths, at a fixed focal plane. With the large FOV of CRISP2, the tilt of the SST focal plane is noticeable and is compensated by tilting the focal plane of the CRISP2 cameras by 1.6\textdegree. The reduced Strehl in the outermost parts of the FOV at short wavelengths is acceptable since CRISP2 targets primarily observations at longer wavelengths. }
  \label{table_CRISP2_strehl}
\end{table}

The large FOV of CRISP2 represents a bigger challenge for its design than was the case with its predecessor CRISP. Integrating the design of the optical system of CRISP2 with that of the secondary optical system of SST, its Schupmann system \citep{2003SPIE.4853..341S}, demonstrated clear differences in image quality between opposite sides of the FOV in one direction. These differences appear because the focal plane of the Schupmann system is tilted by 2.1\textdegree, when measured over a circular FOV with a diameter of  2 arc min, which is a consequence of the asymmetric off-axis design of the Schupmann system. Tilting the focal plane of the cameras (F3) by about 1.6\textdegree compensates the Schupmann focal plane tilt and results in a more uniform image quality over the FOV. In Table \ref{table_CRISP2_strehl} is shown the Strehl values with a tilt in the x-direction, according to the nomenclature of that Table. Without that tilt, the Strehl values at offsets of  (-70",0) are reduced by up to 10\% and at (-60",0) by up to 7\%. 

The need to tilt the cameras of CRISP2 thus is a consequence of the optical design of SST. This built-in feature actually is an advantage, because it provides an opportunity for preventing ghost images and straylight that emanates from reflections between the camera sensor and the reflective cavity of the second (low-resolution) etalon. All camera mountings on CRISP2 therefore allow tilts (rotations) by up to 2\textdegree.

The prescriptions of the optical designs of the CRISP2 FPI system and its wideband systems are given in Appendix \ref{sec:CRISP2}. All lenses are cemented doublets and coated with anti-reflection coatings. The corresponding wideband system uses duplicates of L1 and L4. The present design is for the cameras used with CRISPm, which effectively have 2560$\times$2560 pixels with 5~$\mu$m pixel size, when binned over 2$\times$2 pixels. With an image scale of 0\farcs052 per pixel, images at all wavelengths are critically sampled, or better. The FOV covered with the cameras corresponds to 2.2$\times$2.2 arc min, which is slightly less than needed to capture the entire circular FOV of the etalons. To change cameras in the future, only L4 needs to be replaced.

To enable the best image quality with its large FOV, CRISP2 was made as long as the available space at SST permits, which ultimately led to a design that has a length of 1824~mm, counted from F1 to F3. The CRISP2 wideband system is 452~mm long. The F-ratio at the etalons is "only" 140 in order to maximise its FOV. A consequence of this is that the space available for the etalons is relatively small, and the etalon surfaces that are nearest to the F2 focal plane are at a distance of 70~mm. A distance of 100~mm would have been preferable in order to provide a better defocus of any etalon defects.

On the output end of the F/20 re-imaging system of CRISP2, there is a 25~mm polarising beam splitter. The distance from that to the focal plane is 46.5~mm, which is more than adequate with the cameras used, but that space would not allow very large cameras corner to corner behind the beam splitter. Therefore, placing a polarisation modulator between L4 and the beam splitter would likely not be a possible solution for CRISP2, but should work well with the much longer FPI systems for EST.

The departure from telecentricity at the location of the etalons needs to be small. This departure is a maximum of 0.01\textdegree for CRISP2, which is only about 1/20 of the marginal ray of the F/140 re-imaging system. This has no noticeable effect of the spectral transmission profile or the PSF. Both etalons have external surfaces that are wedged by 0.5--0.6\textdegree relative to the surfaces constituting the cavity, in order to prevent interference fringes involving any of its external surfaces. This wedge angle was set by ray-tracing. 

CRISP2 and its optics is built according to tolerances set by simulations with Zemax. The most critical tolerances are provided in Tables \ref{table_CRISP2_tolerances1} and \ref{table_CRISP2_tolerances2} in Appendix \ref{sec:CRISP2}. These tolerances were considered achievable by lens manufacturers. Other less critical tolerances were adopted from similar design studies of CRISP and CHROMIS. CRISP2 is expected to be installed at SST during the 2025 observing season.
        
\subsubsection{EST-B telecentric optical design.} \label{EST-B_telecentric_design}
EST \citep{2022A&A...666A..21Q} will be equipped with several state-of-the-art science
instruments, including three FPI-based imaging spectropolarimeters, here referred to as EST-B, EST-V, and EST-R. By collecting spectropolarimetric data simultaneously at high cadence, these three instruments will provide unprecedented diagnostics of the thermodynamics, line-of-sight velocities, and the magnetic field of the solar atmosphere, from the deep photosphere to the upper chromosphere. Compared to SST, this will provide a four-fold improvement in spatial resolution and a 16-fold improvement in terms of number of photons detected. Moreover, EST will observe the red and near infrared spectral regions with two FPIs - EST-V and EST-R - covering separately the 500-680~nm and 680--1000~nm wavelength regions, which at SST is covered by a single FPI system, CRISP (soon to be replaced with CRISP2). For example, this allows the full dedication of EST-R to observations of chromospheric magnetic fields, which is very demanding in terms of signal to noise, using the Ca~II 854.2~nm line.  Less challenging measurements of photospheric dynamics and magnetic fields, using various spectral lines, along with chromospheric dynamics in H$\alpha$ can simultaneously be observed with EST-V. The primary role of EST-B is identical to that of CHROMIS, which is spectropolarimetry of the chromosphere in the Ca~II H and K lines, and the spectral resolution of the two systems is the same. These three FPI systems will provide very powerful diagnostics of the entire solar atmosphere, in particular the coupling between the deeper layers and the chromosphere.

Conceptual optical designs of EST-B, EST-V, and EST-R have recently been developed by \citet[][note though that these proposals have not been reviewed by the EST project]{2025arXiv250521053S}, here we summarise their work.  All designs are made with two pixel sizes, either 5~$\mu$m and 12~$\mu$m, or 6.5~$\mu$m and 12~$\mu$m, to cover the range of small pixel sizes with CRISP and CHROMIS, as well as the much larger pixel size of  the DKIST FPI system VTF \citep{2012SPIE.8446E..77K, 2014SPIE.9147E..0ES}. In this publication, we constrain our discussion to the most demanding of the proposed systems for EST, which is EST-B, and we discuss its particular challenges. This system is designed for 5~$\mu$m and 12~$\mu$m pixel sizes, for which the only difference lies in the design of the camera lens (L4).

To provide a context, the layout of EST-B is shown in Fig. \ref{fig:FPI_layouts} and the prescription of  EST-B is given in Appendix \ref{sec:EST-B}, Table \ref{table_EST-B}. As can be seen in Fig. \ref{fig:FPI_layouts}, the overall layout of EST-B is similar to that of CRISP2 and CHROMIS, what differs is primarily the overall scale of EST-B relative to the FPI systems of SST. Another difference is that the lenses of CRISP2 and CHROMIS are cemented, whereas most lenses of EST-B are air spaced, at least in their present designs. The tolerance analysis made \citep[][Appendix D]{2025arXiv250521053S} indicates that the tolerances of EST-B are tighter, but not dramatically so, than those of CHROMIS. 

What makes EST-B (as well as EST-V and EST-R) particularly challenging, when compared to CRISP2 and CHROMIS, are certain aspects of the secondary optical system of EST, POP. The first is that it has a focus curve that must be compensated for when tuning any of the FPI systems to a new wavelength (Sect. \ref{EST_particularities}). The proposal is to compensate that by moving the camera lens, which provides a safe and satisfactory solution that will not have any negative impact on the duty cycle or overall efficiency of the FPI systems \citep{2025arXiv250521053S}. This raises the question of how large focus errors on the input side we must expect and what we can tolerate. We argue, that the focus error on the input side should be very small, about 5~mm, in addition to the variation of the focus curve of POP within the wavelength range of any of the FPI systems. There are two reasons for requiring very small focus errors on the input sides of the FPI systems. The first is simply to minimise the range of movement of the FPI focusing mechanism, which will involve image scale variations and possibly even image quality degradation for large amounts of defocusing. The other reason is to ensure that the focal plane between the FPIs stays away from the surfaces of the two FPIs, such that the FPIs always appear defocused in the focal planes of the science cameras. The problem here is that the movement of the focal plane scales as the square of the focal ratio, such that a 5~mm movement of the F/50 focal plane on the input side of the FPI system leads to movement of 43~mm in the F/147 beam between the etalons. Adding to that a 5~mm focus curve, means an overall movement of the focal plane between the FPIs of nearly 100~mm. This will be acceptable if the FPIs are located at least 200~mm from the nominal focal plane, midways between the two etalons, but obviously very large input focus errors will cause problems.

Considering the overall dimensions and other properties of POP, requiring focus drifts of less than 5~mm seems very demanding. It can be argued, and probably rightfully so, that thermal control of POP plus the EST AO system will be capable of maintaining the focus variation even within such tight limits. However, we prefer a more conservative approach and suggest that each FPI system should have a slow focusing mechanism that involves translating the entire FPI construction, including its wideband system, the polarimeter, the filter wheel and all cameras, along the optical axis.

If such refocusing is made to compensate any (thermal) drifts of POP, it means that the back and forth focusing of the camera lens to compensate the focus curve of POP will not be entirely repetitive, such that for example science data could be recorded with the camera lens in one position, and flats and other calibration data with the camera lens in another position. Or, it could mean that science data at a particular wavelength is recorded with the camera lens in two or more positions, but with flats and calibration data properly cleaning data from artefacts only on parts of the science data. Given the weak polarimetric signals that EST is expected to detect, our concern is that any inconsistencies between the conditions of recording science data and calibration data can lead to weak artefacts. To minimise the risk of such artefacts, \citet{2025arXiv250521053S} have proposed that the FPI systems and POP needs to comply with the following requirements: 
\begin{itemize}
\item Back and forth focusing of the FPI system to compensate the focus curve of POP, by means of moving the camera lens, should be 100\% repetitive during a single observing day to avoid introducing artefacts into the data.
\item The stability of POP should ideally be such that any focus drift during one observing day can be compensated for with the AO system of EST without exhausting its stroke.
\item Any drift of the focal plane, beyond what can be managed by the AO system, should be compensated for by moving the entire mechanical structure holding the FPI system and its wideband system.
\item Whereas the focal plane of POP can be allowed to drift during the day, it is crucial that the stability of POP is sufficiently high that the shape of the focus curve remains the same, such that the focus positions at different wavelengths drift by the same amount. 
\end{itemize}

The above concerns have prompted us to study the imaging performance of the EST FPI systems with a simulated focus curve based on a preliminary design of POP made by Álvaro Pérez García (as of 23 April 2024; private communication). This focus curve shows a variation of 6.3~mm within the wavelength range of EST-B. Based on this focus curve, we have made several optical designs of EST-B, and simulated movement of the camera lens (L4) to compensate the focus curve of POP. From these simulations, we have obtained Strehl values at the optimal focus positions of the camera lens, and we have also measured the change in image scale that such refocusing incurs. This was done for pixel sizes of  5~$\mu$m and 12~$\mu$m, assuming the 6.3~mm focus curve of POP. We have also investigated whether it is possible to modify the design of the camera lens for the 12~$\mu$m pixel size system to reduce the image scale variations of the 12~$\mu$m camera system. The results were the following \citep{2025arXiv250521053S}:
\begin{itemize}
\item Refocusing to compensate the 6.3 mm focus curve of POP with 5~$\mu$m pixel size requires a movement of the camera lens by 1.5~mm and 9.3~mm with the 12~$\mu$m pixel size. This large difference is because the needed movement of the camera lens scales as the square of the magnification (thus, square of the pixel size). 
\item The image scale variation with 5~$\mu$m pixel size, both for the EST-B and relative to that of the wideband system, is about 0.017\%. 
\item With 12~$\mu$m pixel size, the image scale variation for the FPI system is 0.090\% and relative to the wideband system it is 0.038\%, which is significantly larger than with 5~$\mu$m pixel size.
\item Modifying the design of  the 12~$\mu$m camera lens to reduce its image scale variations, gives only a modest improvement in these variations to 0.065\%, and relative to the wideband system to 0.028\%. This variation is still much larger than with 5~$\mu$m pixel size.
\end{itemize}
Table \ref {table_EST-B_Strehls} provides the Strehl values obtained with the 5~$\mu$m pixel camera lens (top), with the 12~$\mu$m pixel camera lens optimised for image quality (middle), and with the 12~$\mu$m pixel camera lens modified to reduce image scale variations (bottom). The camera lens is moved by an amount that is individually set at each wavelength to provide the best compensation for the focus curve of POP. As can be seen, there is a significant price in terms of reduced Strehl for a rather modest reduction of the image scale variations, when refocusing the camera lens. For that reason, we do not recommend a camera lens design that is highly optimised to reduce image scale variations when refocusing.
 
\begin{table}[ht]
 \caption{\label{table_EST-B_Strehls}Strehl values for EST-B connected to POP.}
  \centering
  \small
  \begin{tabular}{rcccccccccc}
        \hline
    \mathstrut
$\lambda~(nm)$&380&400&420&440&460&480&500 \\
FOV dist. \\
    \hline
0.0"&0.97&0.97&0.97&0.97&0.97&0.97&0.98\\
6.0"&0.97&0.98&0.98&0.98&0.98&0.98&0.98\\
11.8"&0.98&0.98&0.98&0.99&0.99&0.99&0.99\\
18.0"&0.97&0.98&0.99&0.99&0.99&0.99&0.99\\
23.6"&0.94&0.96&0.98&0.98&0.99&0.99&0.99\\
30.0"&0.91&0.94&0.95&0.96&0.96&0.96&0.96\\
  \hline
  \hline
 0.0"&0.99&0.99&0.99&0.99&0.99&0.99&1.00 \\
6.0"&0.99&0.99&0.99&0.99&0.99&0.99&1.00\\
11.8"&0.99&0.99&0.99&0.99&0.99&0.99&0.99\\
18.0"&0.99&0.98&0.98&0.98&0.98&0.99&0.99\\
23.6"&0.98&0.97&0.97&0.97&0.97&0.97&0.97\\
30.0"&0.95&0.95&0.95&0.95&0.95&0.96&0.96\\
 \hline
 \hline
0.0"&0.99&0.99&0.98&0.98&0.98&0.97&0.97\\
6.0"&0.99&0.99&0.98&0.98&0.98&0.98&0.98\\
11.8"&0.99&0.99&0.99&0.99&0.99&0.99&0.99\\
18.0"&0.97&0.98&0.98&0.99&0.99&0.99&0.99\\
23.6"&0.94&0.96&0.97&0.97&0.98&0.98&0.99\\
30.0"&0.87&0.90&0.91&0.92&0.93&0.93&0.94\\
 \hline
  \end{tabular}
  \tablefoot{Compensation for the focus curve of POP is made by moving the camera lens of EST-B. Top: 5~$\mu$m pixel size camera lens, middle: 12~$\mu$m pixel size with camera lens optimised for image quality, bottom: 12~$\mu$m pixel size, with camera lens re-optimised to reduce image scale variations.}
 \label{table_EST-B_Strehls}  
\end{table}

To summarise: our efforts to develop compact FPI systems for EST suggest that the challenges they present are not fundamentally different from those of CRISP and CHROMIS, which are highly successful developments. Designing compact FPI systems of less than 4.7~m overall lengths for a  4.2~m telescope, without any folding mirrors and relying entirely on lenses, is entirely possible without sacrifice of image quality. Compensating for the focus curve of POP by moving the camera lens works best with systems designed for cameras with small (5~$\mu$m) pixel size but does not present a problem also with larger (12~$\mu$m) pixel size. Other benefits from choosing a camera with small pixel size is that the stroke of the camera lens movement is smaller, and that a smaller camera lens represents a smaller mass to move. We are therefore not concerned about the need to refocus the EST FPI systems to compensate the focus curve of POP. However,  as discussed already, we believe that it is very important to ensure that the focus error on the input side of the FPI system is very small, about 5~mm or less in addition to that of the focus curve of POP. To achieve that goal, it seems highly advisable to arrange for a slow focusing mechanism of the entire construction holding each FPI system.

\begin{figure}
\center
\includegraphics[angle=0, width=0.9\linewidth,clip]{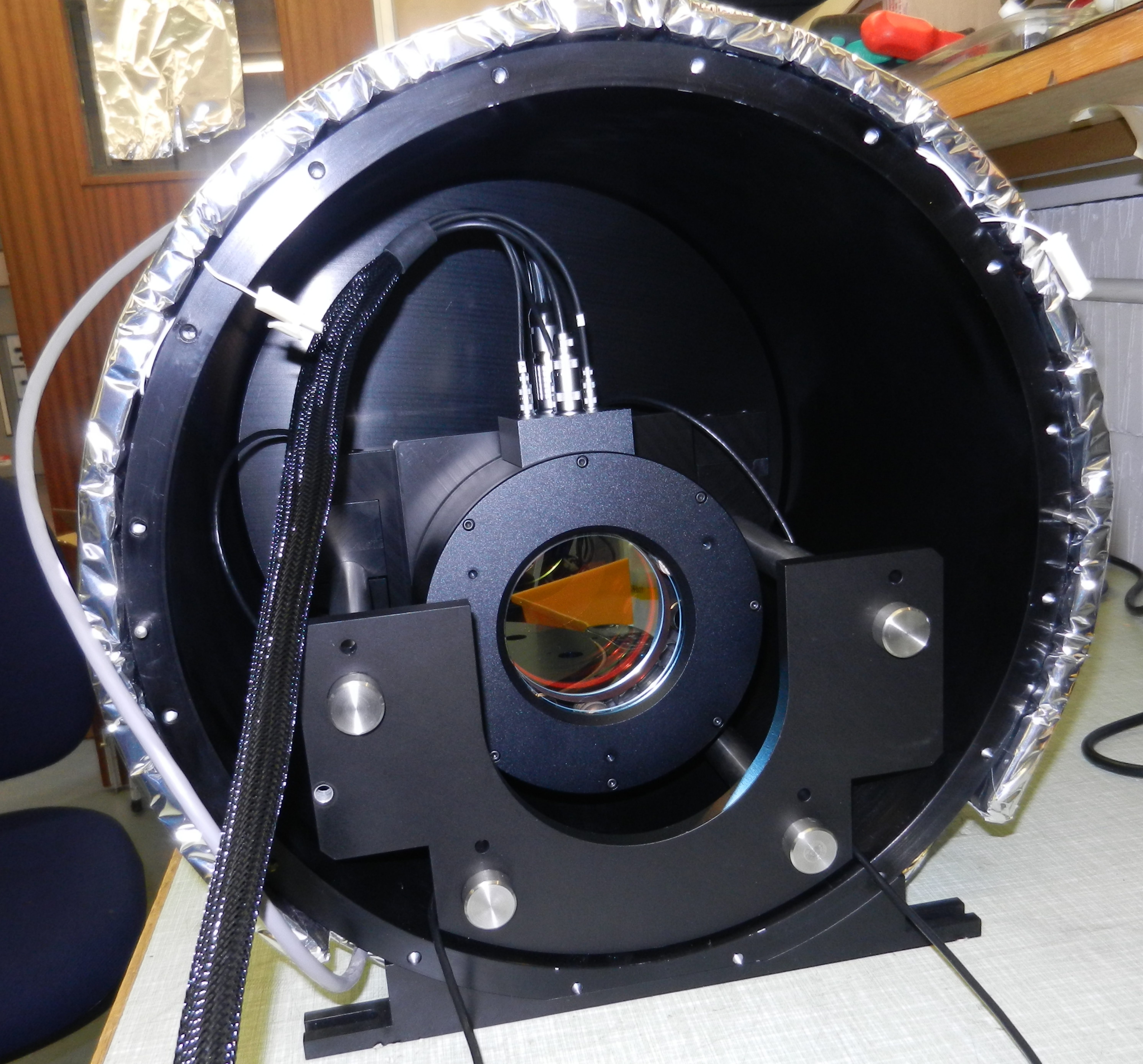}
\includegraphics[angle=0, width=0.9\linewidth,clip]{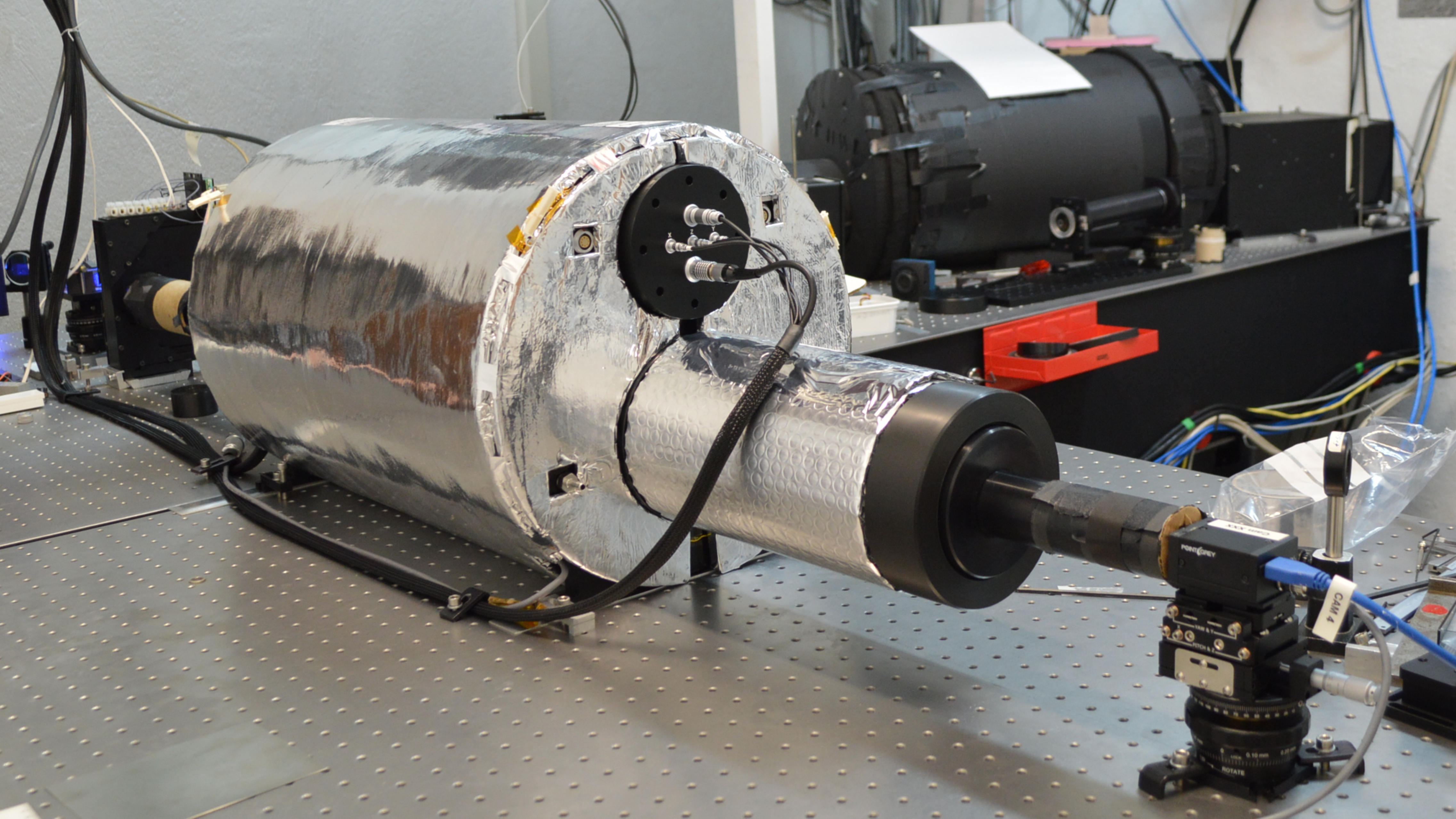}
 \caption{
Exit part of the CHROMIS FPI container during test assembly at ICOS (top) and CHROMIS, fully assembled with one camera, just after mounting on its optical table at SST on La Palma (bottom). The CHROMIS wideband system is not shown. The CRISP FPI container is shown in the upper-right part of the lower panel.}
\label{fig:CHROMIS_pics}
\end{figure}

 \subsection{Mechanical design and enclosure}

The etalons of CRISP, CRISP2 and CHROMIS are enclosed in a container that is sealed by the large lenses L2 and L3, which are mounted with O-rings, thereby eliminating the need for flat windows to enclose the FPIs. This container needs to have substantially larger diameter than the etalons to allow for their mounting and tip-tilt mechanisms. For CHROMIS, with 75 mm clear aperture etalons, the diameter of the container is about 350 mm. The container uses bulkhead feed-through vacuum-proof connections for control of the etalon tuning and the tip-tilt motors that allow alignment of the etalons with respect to the optical axis. This alignment can be made with the entire system sealed and the FPIs fully tunable. Figure \ref{fig:CHROMIS_pics} shows pictures of the FPI container for CHROMIS during test assembly and as mounted on the optical table at SST on La Palma.

After mounting of the etalons in the container, it is flushed with clean nitrogen to provide a favourable environment for the etalons, and the system is then sealed. After tip-tilt alignment of the etalons, power to the tip-tilt motors is switched off permanently. The entire container uses a layer of external insulation to stabilise its temperature against the environment in the optics lab. Temperature control of the container is possible via a heating wire but this has so far not been used. The container is designed and built by IS-Instruments Ltd. A similar container was built and designed for CRISP by the same IS-Instruments team, which was then associated with Hovemere Ltd. We foresee a similar design, but scaled up, for the FPI systems of EST.

\subsection{Control and alignment}
The Fabry-Perot etalons of CRISP2 and CHROMIS and their CS100 controllers were supplied by IC Optical Systems (ICOS). These etalons are piezo tuned and capacitance stabilised to allow active control of the cavity parallelism and scanning via a 16-bit RS232 interface and an HRR16BM interface fitted into the CS100 controller. The temperature induced drift of the system corresponds to a change of cavity spacing of less than 0.1 nm per degree Celsius. With a cavity spacing of 0.787~mm, this corresponds to a wavelength drift of less than 0.08~pm at 630~nm.

The etalons have wedges on their external flat surfaces to eliminate the risk that they cause interference fringes, and are anti-reflection coated for maximum transmission. Pupil stops on x,y stages are mounted between L1 and L2 on the input side and between L3 and L4 on the output sides, and are well centered.

The CHROMIS etalons were pre-assembled inside their FPI container by ICOS and with the L2 and L3 lenses sealing the container for transport to La Palma. Assembling the system in the optics lab on La Palma, and aligning the etalons took less than two days, and the first high-quality data were obtained soon thereafter. Since their installation at SST, the FPI containers of CRISP and CHROMIS have remained closed and sealed.

The alignment of the FPI system is straightforward and is made with lenses mounted and the FPI container sealed. After aligning the entire FPI system with the optical axis of the telescope, the tip-tilt motors of the FPIs are used to ensure that the cavity of the first (high-resolution) etalon is precisely perpendicular to this optical axis, and then the tip-tilt angle of the low-resolution etalon is adjusted such that the second pupil stop rejects all ghost images, which corresponds to a tip-tilt of $1/(2 F\#)$ radians, where $F\#$ is the focal ratio of the beam at the etalons. 

\section{Cameras and polarimeters for CRISP2 and CHROMIS}
\subsection{Cameras}
Both CRISPm and CHROMIS operate with cameras that have small pixels, effectively 5-5.5~$\mu$m. The relatively small full well of these cameras is not a limitation, since we use short integration times (less than 15~ms) in order to  maximise the quality of the MOMFBD restored images. Additionally, the choice of smaller pixel sizes gives access to a much larger selection of cameras and sensors than is the case with cameras based on sensors with large pixels. 

\subsubsection{CRISPm and CRISP2 cameras}
Initially, CRISP used four CAM1M100 back illuminated CCDs delivered by the Sarnoff Corporation in 2008. These were 1024$\times$1024 pixel cameras having 16$\mu$m pixel size and a read-out time of 10~ms. Readout noise was approximately 25~e$^-$, full well about 100~ke$^-$. The cameras were exposed with the aid of a chopper that typically delivered 37 frames per second, with 17~ms exposure and 10~ms readout. The quantum efficiency (QE) reaches up to 65\% around 600~nm and 45\% at 850~nm. The main limitations of this camera are the number of pixels, which does not match the FOV of CRISP, the duty cycle that is only 63\%, and the back scattered light at 854~nm, caused by the semi transparent silicon layer of the back illuminated sensor. The latter required processing the data with quite elaborate calibration and compensation for stray light \citep[][their Appendix]{2013A&A...556A.115D}.

The presently used cameras with CRISPm (identical to CRISP but with a new camera lens), and that will also be used with CRISP2, were installed in 2022. These are the MX262RG-GP-X8G3-MTP-LA cameras from Ximea, based on the Gpixel GMAX0505RF sensor. This near-infrared enhanced sensor has  5120$\times$5120 2.5~$\mu$m pixels that are binned to effectively deliver 2560$\times$2560 5~$\mu$m pixels, which is sufficient to cover the FOV of CRISP2. The sensor has global shutter, which is necessary for polarimetry, 100\% duty cycle, a QE that peaks at about 68\% and that reaches about 35\% at 850~nm, and a read noise of 2.6e$^-$. The lower QE at 850~nm, compared to that of the Sarnoff camera, is more than well compensated for by the 100\% duty cycle and the absence of back scatter. The full well of the binned pixels is 26~ke$^-$ but we avoid exposing these cameras to more than about 19~ke$^-$. We typically run these cameras with 12.5~ms integration times and 80~Hz frame rate. The Ximea cameras were delivered without cover glass on the sensor in order to minimise the risk of interference fringes. We have four such cameras plus one spare permanently installed on CRISPm, and soon CRISP2. The images shown in Fig. \ref{fig:AR+plage} were recorded with these cameras. 

The most obvious improvement of the new cameras, when compared to the previous Sarnoff CCDs, is the much larger FOV that can be covered. Another important improvement is the possibility to shorten the integration time without reduction of the duty cycle: the Sarnoff cameras were used at 63\% duty cycle with 17~ms, the new cameras are used at 100\% duty cycle with 12~ms integration. This more than doubles the number of frames per second, which together with the use of shorter exposure times significantly improves the image quality of the MOMFBD reduced images. Another clear improvement is the quality of images at 854~nm, because of the absence of back scattered light that degraded the Sarnoff images. This back scattered light contribution amounted to about 30\%  of the detected photons, and its removal in software is far from trivial and furthermore leaves the noise from the back scattered photons uncompensated for. The third improvement is the read noise, which is reduced by almost a factor 10 with the new cameras. This should be of particular importance when observing dark sunspots and outside the solar limb, in particular when observed in the core of a strong spectral line. However, a direct comparison between data recorded of such "dark" targets with the Sarnoff cameras and the new Ximea cameras has not yet been attempted.

\subsubsection{CHROMIS cameras}\label{CHROMIS_cameras}
The camera initially installed with CHROMIS, and used until the end of the 2024 observing season, is the Point Grey Grasshopper3 GS3-U3-23S6M camera, which is based on the Sony IMX174 sensor with 1920$\times$1200 5.86~$\mu$m pixels. These low-cost cameras delivered excellent data but the number of pixels is too small to cover the FOV of CHROMIS. Furthermore, more recent generations of the Sony Pregius sensors have higher QE at short wavelengths and lower read noise. The CHROMIS images shown in Fig. \ref{fig:AR+plage} were recorded with this camera.

The cameras now (May 2025) being installed for CHROMIS are the MX203MG-SY-X4G3-FF from Ximea. This uses the IMX531 sensor from Sony with 2.74~$\mu$m pixels that are binned on-chip to 2256$\times$2256 5.48~$\mu$m pixels. The read noise is about 1.4~e$^-$. The primary advantages of this camera is the global shutter, the 100\% duty cycle, high frame rate (up to 110~Hz) and the very high QE of more than 70\% at 390~nm. The cameras are delivered without cover glass to reduce the risk of interference fringes. The similar pixel size of this and the earlier Ptgrey camera eliminates the need to replace the camera lens of CHROMIS.

\subsection{Polarimeters}
The polarimeter for CRISP, which is identical to that of CRISPm and CRISP2, and CHROMIS are of different designs. The CRISP polarimeter \citep{2021AJ....161...89D}, which targets the entire 500-900 nm wavelength range, is based on the use of two Ferroelectric liquid crystals (FLCs) that have constant retardance but that switch their fast axis orientation between two states separated by a switching angle, which is typically 45\textdegree. The preference for FLCs, rather than liquid crystal variable retarders, is because of their fast switching speeds, which is essential at the frame rates of the CRISPm and CRISP2 cameras (approximately 80 Hz). To cover the entire wavelength range of CRISPm, the design strategy used is to optimise the efficiency, rather than the achromaticity, of the polarimetric response of the modulator, as first proposed by \citet{2010ApOpt..49.3580T}. Simulations were used to optimise the retardances and angles of the two FLCs, as well as their sensitivity to various manufacturing and alignment errors and temperature changes. The finally assembled modulator, which is about 57 mm thick, of which 42 mm is glass, was characterised using a facility lab spectropolarimeter at High Altitude Observatory, and the modulation efficiencies were measured at several wavelengths from 517 nm to 854 nm. The overall efficiency of Stokes Q, U, and V measurements was found to peak at over 96\% at 617–630 nm and to drop to about 90\% at 854 nm, with smoothly varying and well balanced efficiencies for all Stokes parameters at all wavelengths. It is temperature controlled with a stability of ±0.2°. The polarimeter has been in use with CRISP since 2015 and is located close to the input focal plane of CRISP. Figure \ref{fig:LOS_magnetic_field} shows examples of  photospheric and chromospheric LOS magnetic field maps obtained with CRISP and this polarimeter.

For the polarimeter for CHROMIS, FLCs were not used because of their limited transparency at the wavelengths of the Ca II H and K lines around 390 nm.
Instead, the design uses two LCVRs with clear apertures of 30 mm between high-quality optical windows of fused silica with dimensions similar to those of the CRISP polarisation modulator. The LCVRs were made by Meadowlark using LC material that allows for a tuning settling time of a few ms, much faster than the hundreds of milliseconds typically required to tune an LCVR and only marginally slower than the typical 0.1~ms switching time of FLCs. To reach fast tuning speed the optics must be held at an elevated temperature of 50\textdegree C. While fast, these LCVRs are not available with large retardance, which presents a difficulty in designing a modulator that can achieve any modulation scheme. It is desirable to have the shortest possible modulation cycle to suppress the effects of seeing-induced crosstalk \citep{2012ApJ...757...45C} which would call for a 4-state scheme to recover the 4 elements of the Stokes vector. At the same time, 6-state schemes with individual measurements of Q, U and V that implement beam switching \citep[e.g.,][]{1993A&A...278..231S, 1998A&A...331..760B} reduce polarisation artefacts even further and also allow polarisation measurements with close to 100\% efficiency dedicated to Stokes V measurements. We believe this to be of particular importance when observing weak quiet Sun chromospheric magnetic fields, for which Q and U signals are expected to be negligibly small. The straightforward solution is to build a "universal modulator" that can rotate any Stokes vector to any other Stokes vector, e.g., using the combination of an LCVR tunable over a full wave of retardance aligned to the polarization analyser reference followed by an LCVR tunable over a half wave of retardance aligned at 45\textdegree. Such a modulator can achieve any modulation scheme by selecting appropriate drive voltages for each state. However, these fast LCVRs can reach at most half-wave retardance for the CHROMIS wavelength range. The CHROMIS modulator therefore uses two LCVRs tunable between 0 and 1/2 wave, with the first oriented at 22.5\textdegree relative to the polarization analyser reference, and the second at 45\textdegree. This setup allows both a 6-state "Stokes definition" scheme as well as a balanced 4-state scheme. This polarimeter, which was recently (May 2025) installed at SST, will be described in detail in a forthcoming paper (De Wijn et al., in prep.).

\section{Conclusions}\label{conclusions}
We have for the first time presented the rationale and some details of the designs of two of the most productive science instruments in ground-based solar physics, CRISP and CHROMIS. These are imaging spectropolarimeters based on dual FPIs embedded within their telecentric reimaging systems. Both represent state of the art in narrowband imaging and polarimetry of the solar surface. CHROMIS is to our knowledge the first FPI based tunable narrowband filter system capable of observing the Sun at wavelengths below 400~nm. We also note that, after the actual submission of this paper, CRISP2 was successfully installed at SST in July 2025, and is performing according to specifications (a paper describing its characteristics and the calibration procedures used with CRISP and CRISP2 is in preparation by de la Cruz Rodr{\'\i}guez et al.).

We use numerical simulations \citep{2006A&A...447.1111S} to quantify the damaging effects of realistic levels of cavity errors on the overall transmission and the fidelity of the spectral transmission profile of CRISP and CHROMIS, and show how to mitigate this by combining a high-resolution high-reflectivity etalon with a low-resolution etalon with much lower reflectivity (Sects. \ref{simulation_insights} and \ref{simulations}). We also use simulations to calculate the degradation of the spatial PSF by multiple reflections in the etalon cavities, and the improvement in Strehl achieved by optimum focus compensation \citep[first described by][]{2006A&A...447.1111S}. Given a spectral resolution of the system, and a target Strehl (95\%), this establishes the minimum focal ratio of the re-imaging system at the location of the etalons. This in turn determines the maximum diameter of the FOV with given clear apertures of the etalons, or their needed minimum clear aperture diameter, if the FOV diameter is given. 

The FPI parameters of CRISP, CRISP2 and CHROMIS have been optimised with the above approach. The same strategy has been applied to the conceptual design of three FPI systems with a FOV diameter of 1 arc min \citep{2025arXiv250521053S}, for the 4.2~m European Solar Telescope \citep[EST;][]{2022A&A...666A..21Q}. The result is that EST needs FPI clear aperture diameters of 180~mm for the two FPIs covering the 500-1000~nm wavelength region, and 135~mm for the 380-500~nm FPI system because  of its lower spectral resolution. These FPI diameters are significantly smaller than the 250~mm clear apertures of the VTF FPI system \citep{2012SPIE.8446E..77K, 2014SPIE.9147E..0ES} developed for the 4~m Daniel K. Inouye Solar Telescope \citep[DKIST;][]{2018SPIE10700E..0VW,2020SoPh..295..172R}.

We also describe the optical design of CRISP, CRISP2 and CHROMIS, the details of which are consequences of the following:
\begin{itemize}
\item The FPI clear aperture diameter and focal ratio of its imaging system, obtained from numerical simulations (Sect.  \ref{simulations}) 
\item The decision to use telecentric reimaging rather than a collimated mount (Sect. \ref{telecentric_vs_collimated})
\item The decision to use lenses instead of mirrors for reimaging.
\item The decision to build the FPI system around a compact straight-through optical system without folding mirrors (Sect. \ref{design_aspects}).
\end{itemize}

We demonstrate that compact optical designs of CRISP, CHROMIS and CRISP2 are feasible with high image quality (Strehl close to 95\% or higher). In a previous publication, we have verified the high image quality of CRISP by comparing the granulation contrast measured through CRISP with that through a wideband filter centered on the same wavelength \citep[Sect. \ref{CRISP_performance} and ][]{2019A&A...626A..55S}. We give here further examples of high-quality data recorded with CRISP and CHROMIS  (Sect. \ref{CRISP_performance}). These data represent state of the art in narrowband imaging and polarimetry of the solar surface. We emphasise the crucial importance of, and synergy with, the wideband system supporting all processing of the narrowband data, and describe the methods used to calibrate and process the data (Sect. \ref{image_restoration}).

The same tools and rules used to design the optical systems of  CRISP, CHROMIS and CRISP2 have also been employed to develop the conceptual optical designs of three FPI-systems \citep{2025arXiv250521053S} for the future 4.2~m EST. These are straight-through compact optical systems without folding mirrors and total lengths of less than 4.7~m. Here, we emphasize the similarities between the FPI systems of SST and those proposed for EST, but also draw attention to potential challenges associated with any drifts of  the EST secondary optical system POP, which precedes the FPI systems. Nonetheless, we conclude that building FPI systems for EST to a large extent presents the same challenges as those of CRISP and CHROMIS. Making these systems compact, both in terms of their lengths and FPI diameters, should offer several advantages, including manufacture, alignment, stability, flexibility in changes of image scale, and costs. These design aspects make the FPI systems both robust and highly performing.

We finally note, that the optical design of the EST FPI systems described here, which is in stark contrast to that of the Visible Tunable Filter system \citep[VTF; ][]{2012SPIE.8446E..77K, 2014SPIE.9147E..0ES} for DKIST, primarily represent proposals to demonstrate their feasibility but likely do not represent final designs. We emphasise that our design of FPI systems for EST is based on a provisional optical design of POP that is likely to undergo modifications and improvements. We thank the EST Project Office for making this information available before publication.

\begin{acknowledgements}
The Swedish 1-m Solar Telescope is operated on the island of La Palma by the Institute for Solar Physics of
Stockholm University in the Spanish Observatorio del Roque de los Muchachos of the Instituto de Astrofísica
de Canarias. The Institute for Solar Physics is supported by a grant for research infrastructures of national importance from the Swedish Research Council (registration number 2021-00169).  This work also has received funding from the European Union’s Horizon 2020 research and innovation programme under grant agreement
No 824135.

The European Solar Telescope project is supported by a grant for research infrastructures from the Swedish Research Council (registration number 2023-00169).  

CRISP and CHROMIS were funded by the Wallenberg Foundations, registration numbers 2003.0037 and 2012.1005. 

J.de la Cruz Rodr\'iguez gratefully acknowledges funding by the European Union through the European Research Council (ERC) under the Horizon Europe program (MAGHEAT, grant agreement 101088184). 

G. Scharmer acknowledges valuable provisional information related to EST, POP and requirements for the FPI systems by the EST Science Advisory Group (SAG) and the EST Project Office (in particular Claudia Ruiz de Galarreta) and discussions with and feedback from TIS consortium members Javier Bailén, Luis Bellot Rubio, Luca Giovanelli, Matteo Munari and Javier Sanchez. Álvaro Pérez García is thanked for providing details of the design of POP.

\end{acknowledgements}

\begin{appendix}
\section{\label{table_CRISP2}CRISP2 telecentric design}
Table \ref{table_CRISP2} shows the design of the narrowband CRISP2 system and Table \ref{table_CRISP2_WB} its wideband system. The system allows a simple change of image scale by replacing L4 (the camera lens) and the tube holding that lens from pupil stop P2. The pupil stops are made with 10\% larger diameters than their corresponding pupil diameters and are mounted on (x,y) stages for fine tuning of  their positioning on the pupil image.
\label{sec:CRISP2}

\begin{table}[h]
\caption{Prescription of the CRISP2 FPI system.}
  \centering
  \small
   \setlength{\tabcolsep}{2.8pt}
  \begin{tabular}{rrrrrcc}
    \hline
    \mathstrut
No. & Radius & Thickness & Lens dia& Beam dia & Glass & Label \\
& (mm) & (mm) & (mm) & (mm) & & \\
    \hline
0 & & 163.46 & & & & F1 \\
1 & 111.91 & 4 & 50 & 39 & N-LASF45 & L1 \\
2 & 51.02 & 13 & 50 & & N-PSK53A &  L1 \\
3 & -175.50 &145.86 & 50 & & & L1 \\
4 & & 418.26 & & 3.15* & &  P1 \\
5 & 342.59 & 8 & 120 && N-SF2 & L2 \\
6 & 141.56 &  26 & 120 &&N-BK7 & L2  \\
7 & -370.98 & 60 & 120 & 100 && L2 \\
8  & & 35 & 105 & 99.4 & Silica & FPI1 \\
9 & & 0 & 105 & & & FPI1 \\
10 & & 25 & 105 & & Silica & FPI1 \\
11 &.&  0.79 & 105 & && Cavity \\
12 & & 35 & 105 & & Silica & FPI1 \\
13 & & 70 & 105 & && FPI1 \\
-&&&&&& F2 \\
14 & & 70 &105 & 98.36 && FPI2 \\
15 & & 35 &105 && Silica & FPI2 \\
16 & & 0.30 & 105 &&& Cavity \\
17 & & 25 & 105 && Silica & FPI2 \\
18 & & 0 & 105 &&& FPI 2 \\
19 & & 35 & 105 && Silica & FPI2 \\
20 & & 60 & 105 & 98.9 & & FPI2 \\
21 & 370.98 & 26 & 120 & 100 & N-BK7 & L3 \\
22 & -141.56 & 8 & 120 && N-SF2 & L3 \\
23 & -342.59 & 418.35 & 120 &&&  L3 \\
24 & & 59.22 && 3.36* &&  P2 \\
25 & 91.00 & 3 & 26 && N-SF57 & L4 \\
26 &30.26 & 7 &26 &&N-LAF2 & L4 \\
27 &-71.24 & 1 & 26 & 16 &&L4 \\
28 && 25 &&& SILICA & PBS \\
29 && 46.46 &&&&  PBS \\
30 & &&&&&F3 \\
 \hline
 \end{tabular}
  \tablefoot{F1-F3 are focal planes, L1-L4 are cemented doublet lenses, P1-P2 are pupil planes, and PBS is the polarising beam splitter. The total length from F1 to F3 is 1824 mm. *The actual pupil diameter is given in the Table - the pupil stops should be about 10\% larger. }
\end{table}

\begin{table}[h]
\caption{Prescription of the CRISP2 wideband system.}
  \centering
  \small
     \setlength{\tabcolsep}{2.8pt}
  \begin{tabular}{rrrrrcc}
        \hline
    \mathstrut
No. & Radius & Thickness & Lens dia& Beam dia & Glass & Label \\
& (mm) & (mm) & (mm) & (mm) & & \\
    \hline
1 && 155.50&&&& F1 \\
2 & 111.91 & 4 & 50 && N-LASF45 & L1 \\
3 & 51.02 & 13 & 50 && N-PSK53A & L1 \\
4 & -175.5 & 145.89 &50 &&& L1 \\
5 && 59.08 && 3.15* && P1/P2 \\
6 & 91.00 & 3 & 26 && S-SF57 & L4 \\
7 & 30.26 & 7 & 26 && N-LAF2 & L4 \\
8 & -71.24 & 1 & 26 && L4 \\
9 && 25 &&& N-BK7 & PDBS \\
10 && 38.65 \\
11 &&&&&&F3 \\
 \hline
  \end{tabular}
  \tablefoot{Labelling is made to agree with that of CRISP2 FPI system, the prescription of which is shown in Table \ref{table_CRISP2}. F1 and F3 are focal planes, L1 and L4 are cemented doublet lenses that are duplicates of the corresponding lenses in the CRISP2 FPI system, P1 is a pupil plane, and PDBS is the phase diversity beam splitter. *This is the pupil diameter - the pupil stop should be about 10\% larger. The total length from F1 to F3 is 452 mm. The distance from the first surface of L1 to the last surface of L4 is 233~mm.}
  \label{table_CRISP2_WB}
\end{table}

\subsection{CRISP2 Tolerances}
Tables \ref{table_CRISP2_tolerances1} and \ref {table_CRISP2_tolerances2} show the most critical tolerances on the manufacture of the lenses and their assembly. As regards these tolerances, it should be emphasised that these are of two types: the first concerns the manufacture and mounting of lens assemblies, such as the L1--L4 doublets. The high requirement on image quality drives tight tolerances on curvatures, wedges and centerings of the individual that are in several cases demanding. However, modern lens making can handle the manufacture and assembly of much more complex designs than those of the FPI systems, and checking some of our tolerances with tentative contractors have not given any reasons for caution.

The other part of the tolerances concern the installation and centering of the lenses relative to each other within the mechanical structure holding the FPIs and the lenses L1-L4. Here, tolerances on centering are typically on the order of 0.5 mm, which is not demanding. The tolerance on tilt ($\pm 0.05$\textdegree) is much more demanding but fully feasible, given some help tools, and was in fact achieved with CHROMIS.

\begin{table}[ht]
\caption{\label{table_CRISP2_tolerances1}Optical tolerances of the CRISP2 doublet lenses. }
  \centering
  \small
  \begin{tabular}{ll}
        \hline
    \mathstrut
Parameter & Tolerance \\
\hline
Radius error & $\pm$0.1\% \\
Thickness & $\pm$0.1 mm \\
Form error & $\pm$1/3 wavelength \\
Wedge & $\pm$30 seconds of arc \\
Index error & $\pm$0.0002 * \\
Abbe error  & +-0.3\% * \\
 \hline
  \end{tabular}
  \tablefoot{*The index of the glass is measured after purchase, and the system is re-optimised with the measured glass data.}
\end{table}

\begin{table}[h]
 \caption{\label{table_CRISP2_tolerances2}Summary of CRISP2 assembly tolerances.}
  \centering
  \small
    \begin{tabular}{l|l}
        \hline
    \mathstrut
Parameter & Tolerance \\
\hline
Decentration of beam expander \\
relative to the beam reducer: & $\pm$0.5~mm* \\
Tilt of beam expander \\
relative to the beam reducer & $\pm$0.05 degrees ** \\
Decentration of beam expander: \\
relative to the telescope: & $\pm$ 0.5~mm * \\
Tilt of beam expander \\
relative to the telescope: & $\pm$0\textdegree.05 ** \\
Decentration of achromats \\
relative to the mounting flange: & $\pm$0.1~mm ** \\
Tilt of the achromats \\
relative to mounting flange: & $\pm$0.05 degrees \\
Distance between small \\
and large achromat: & $\pm$0.25~mm \\
Length of FP box: & $\pm$1mm \\
 \hline
  \end{tabular}
  \tablefoot{* critical tolerances. ** highly critical tolerances}
\end{table}

\clearpage

\section{EST-B telecentric design}
\begin{figure}[h]
\center
\includegraphics[angle=0, width=0.99\linewidth,clip]{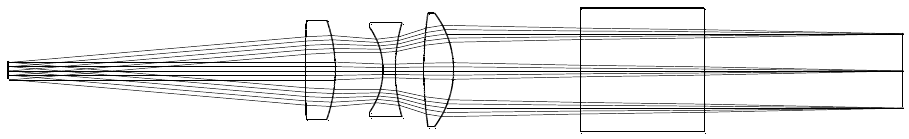}
\includegraphics[angle=0, width=0.99\linewidth,clip]{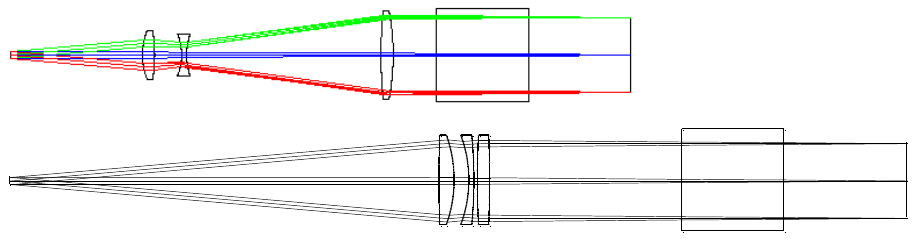}
 \caption{
Layout of three camera lenses designed for EST-B narrowband FPI system. Top: camera lens designed for 5~$\mu$m pixel size, middle: 12~$\mu$m pixel size camera lens optimised for image quality, bottom: 12~$\mu$m pixel size camera lens optimised for reduced image scale changes when refocusing. The 5~$\mu$m and 12~$\mu$m systems are drawn at different scales. For further details about the systems, see Tables \ref{table_EST-B_5my_camera_lens} and \ref{table_EST-B_12my_camera_lens1}.}
\label{fig:EST-B_camera_lenses}
\end{figure}

Table \ref{table_EST-B} shows the optical design of the fixed part of the optical system for EST-B, from L1 to P2, together with an idealised model of the telescope and POP. In Tables \ref{table_EST-B_5my_camera_lens}--\ref{table_EST-B_12my_camera_lens2}, we show the prescriptions of the optical design of three different camera lenses for EST-B, the layouts of which are shown in Fig. \ref{fig:EST-B_camera_lenses}. The first of these designs is of a camera lens for 5~$\mu$m pixel size, the remaining two are for cameras with 12~$\mu$m pixel size, one for which image quality is prioritised (Table \ref{table_EST-B_12my_camera_lens1}), the other prioritising reduced image scale variations when the camera lens is refocused (Table \ref{table_EST-B_12my_camera_lens2}).
\label{sec:EST-B}

\begin{table}[h]
 \caption{\label{table_EST-B}Prescription of the EST-B FPI "mother" system.}
  \centering
  \small
  \setlength{\tabcolsep}{2pt}
  \begin{tabular}{rrrrrcc}
    \hline
    \mathstrut
No. & Radius & Thickness & Lens dia & Beam dia & Glass & Label \\
& (mm) & (mm) & (mm) & (mm) & & \\
    \hline
 --&&&&&& \bf{Telescope} \\
 --&&&&&& \bf{+ POP} \\
0 && infinity&&&& \\
1 && 100 &&&& System stop \\ 
2 && 100 &&&& Ideal lens \\
-- &&&&&& \bf{FPI system} \\
3 && 100 &&&& F1 \\
4 && 80 && 64 & Silica & WBBS \\
5 && 195.33 && 64 \\
6 & 284.96 & 8 & 82 & 70 & PBL1Y & L1 \\
7 & 117.84 & 14 & 82 & 70 & N-FK51A & L1 \\
8 & -340.93 & 345.96 & 82 & 70 \\
9 && 800.82 && 7.9 && P1 \\
10 & 790.98 & 24 & 160 & 147 & N-FK5 ** & L2 \\
11 & -292.40 & 39.89 & 160 & 147 \\
12 & -253.37 & 16 & 150 & 142 & PBM2Y & L2 \\
13 & -463.09 & 407.93 & 160 & 142 \\
14 && 140 & 160 &139 & Silica & FPI1 \\
15 && 200 & 160 &139 \\
16 && 200 && 136 && F2 \\
17 && 140 & 160 & 139 & Silica &FPI2 \\
18 && 407.93 & 160 & 139 \\
19 & 463.09 & 16 & 160 & 143 & PBM2Y & L3 \\
20 & 253.37 & 39.89 & 150 & 143 \\
21 & 292.40 & 24 & 160 & 147 & N-FK5 ** & L3 \\
22 & -790.98 &  800.25 & 160 & 147 \\
23 &&&&&& P2 \\
 \hline
  \end{tabular}
  \tablefoot{The camera lens is not included here. F1-F3 are focal planes, L1-L4 are cemented doublet lenses, P1-P2 are pupil planes, and WBBS is the wideband beam splitter. The total length from F1 to F3 is 4400 mm.}
\end{table}

\begin{table}[h]
\caption{ \label{table_EST-B_5my_camera_lens}The 5~$\mu$m pixel size camera lens for EST-B narrowband.}
  \centering
  \small
  \setlength{\tabcolsep}{3pt}
  \begin{tabular}{rrrrrcc}
    \hline
    \mathstrut
No. & Radius & Thickness & Lens dia & Beam dia & Glass & Label \\
& (mm) & (mm) & (mm) & (mm) & & \\
    \hline
23&&120.15&&8&&P2\\
24&850.56&12&40&29&S-BSM16&L4\\
25&-60.34&19.28&40&29&&L4\\
26&-29.73&5&34&28&PBL6Y&L4\\
27&69.083&11.50&38&28&&L4\\
28&158.60&12&46&37&N-FK51&L4\\
29&-38.15&51.22&46&37&&L4\\
30&&50&&35&SILICA&PBS\\
31&&80&&35&&F3\\
 \hline
  \end{tabular}
  \tablefoot{P2 is the second pupil stop, F3 the final focal planes, L4 is an air spaced triplet lens, and PBBS is the polarising beam splitter.}
\end{table}

\begin{table}[h]
\caption{\label{table_EST-B_12my_camera_lens}The 12~$\mu$m pixel size camera lens for EST-B narrowband.}
  \centering
  \small
  \setlength{\tabcolsep}{3pt}
  \begin{tabular}{rrrrrcc}
    \hline
    \mathstrut
No. & Radius & Thickness & Lens dia & Beam dia & Glass & Label \\
& (mm) & (mm) & (mm) & (mm) & & \\
    \hline
23&&128.881&&8&&P2\\
24&62.760&12&48&30&N-PK51&L4\\
25&-141.772&26.368&48&30&&L4\\
26&-55.048&5&40&24&PBL1Y&L4\\
27&63.972&190.494&40&24&&L4\\
28&412.253&12&86&78&N-PK51&L4\\
29&-223.835&42.417&86&78&&L4\\
30&&90&&77&SILICA&PBS\\
31&&100&&77&&F3\\
 \hline
  \end{tabular}
  \tablefoot{The camera lens is optimised for image quality.  P2 is the second pupil stop, F3 the final focal planes, L4 is an air spaced triplet lens,, and PBBS is the polarising beam splitter.}
  \label{table_EST-B_12my_camera_lens1}
\end{table}

\begin{table}[h]
\caption{\label{table_EST-B_12my_camera_lens2}The 12~$\mu$m pixel size camera lens for EST-B narrowband.}
  \centering
  \small
  \setlength{\tabcolsep}{3pt}
  \begin{tabular}{rrrrrcc}
    \hline
    \mathstrut
No. & Radius & Thickness & Lens dia & Beam dia & Glass & Label \\
& (mm) & (mm) & (mm) & (mm) & & \\
    \hline
23&&420.74&&&&P2\\
24&1248.34&15&88&&N-FK51A&L4\\
25&-130.41&14.48&88&&&L4\\
26&-124.35&5&88&&PBL1Y&L4\\
27&-419.51&3.15&88&&&L4\\
28&736.58&12&88&&N-SK5HTI&L4\\
29&-1328.11&188.27&88&&&L4\\
30&&100&&&SILICA&PBS\\
31&&120&&&&F3\\
 \hline
  \end{tabular}
  \tablefoot{The camera lens is optimised to reduce image scale variations.  P2 is the second pupil stop, F3 the final focal planes, L4 is an air spaced triplet lens,, and PBBS is the polarising beam splitter.}
\end{table}

\subsection{EST-B Tolerances}
Tables \ref{table_EST-B_tolerances1} and \ref {table_EST-B_tolerances2} show the most critical tolerances relating to the manufacture of the lenses and their assembly for EST-B. These tolerances are of two types: the first concerns the manufacture and mounting of lens assemblies, such as the L1, L2 and L3 doublets. The high requirement on image quality drives tight tolerances on curvatures, wedges and centerings of the individual that are in several cases demanding. However, modern lens making can handle the manufacture and assembly of much more complex designs than those of L1-L3 for the EST-B FPI system, and checking some of our tolerances with tentative contractors have not given any reasons for concern. The tolerances on L4, which is an air spaced triplet, are particularly tight, and to meet these tolerances we have outlined a special step-by-step procedure for its mounting, which is described in Appendix D of \cite{2025arXiv250521053S}.

The other part of the tolerances concern the installation and centering of the lenses relative to each other within the mechanical structure holding the FPIs and the lenses L1-L4. Here, tolerances on centering are typically on the order of 0.5 mm, which is not demanding. The tolerance on tilt ($\pm 0.05$\textdegree) is much more demanding but fully feasible given some help tools, and was in fact achieved with CHROMIS.

\begin{table}[h]
 \caption{\label{table_EST-B_tolerances1} Optical tolerances of the EST-B lenses L1 and L2/L3.}
  \centering
  \small
  \begin{tabular}{ll}
        \hline
    \mathstrut
Parameter & Tolerance \\
\hline
Radius error & $\pm$0.1\% \\
Thickness & $\pm$0.1 mm \\
Wedge angle L2/L3& $\pm$ 0.003\textdegree \\
 Wedge angle L1 & $\pm$0.008\textdegree   \\
 \hline
  \end{tabular}
  \end{table}

\begin{table}[h]
 \caption{\label{table_EST-B_tolerances2}Assembly tolerances for lenses L1-L3.}
  \centering
  \small
  \begin{tabular}{ll}
        \hline
    \mathstrut
Parameter & Tolerance \\
\hline
Decenter of entire FPI system relative to POP & $\pm$0.5 mm \\
Decenter of  L3+L4 relative to L1+L2 & $\pm$0.5 mm \\
Tilt of entire FPI system relative to POP & $\pm$ 0.05 degrees \\
Tilt of camera L3+L4 relative to L1+L2 & $\pm$ 0.05 degrees \\
Decenter between individual lenses & $\pm$ 0.1 mm \\
Tilt of individual lenses & $\pm$ 0.05 degrees \\ 
 \hline
  \end{tabular}
 \end{table}

\end{appendix}
\end{document}